\documentclass[10pt,preprint] {aastex}

\usepackage{graphicx,epsfig,natbib,color}
\usepackage{lscape}
\usepackage{graphicx}
\usepackage{epsfig}
\usepackage{natbib}

\begin{document}
\title{Transition from circular-ribbon to parallel-ribbon flares associated with a bifurcated magnetic flux rope}

\author{Z. Zhong\altaffilmark{1,2}, Y. Guo\altaffilmark{1,2}, M. D. Ding\altaffilmark{1,2}, C. Fang\altaffilmark{1,2}, Q. Hao\altaffilmark{1,2}}

\affil{$^1$ School of Astronomy and Space Science, Nanjing University, Nanjing 210023, China}
\email{guoyang@nju.edu.cn, dmd@nju.edu.cn}
\affil{$^2$ Key Laboratory for Modern Astronomy and Astrophysics (Nanjing University), Ministry of Education, Nanjing 210023, China}

\begin{abstract}
Magnetic flux ropes play a key role in triggering solar flares in the solar atmosphere. In this paper, we investigate the evolution of active region NOAA 12268 within 36 hours from 2015 January 29 to 30, during which a flux rope was formed and three M-class and three C-class flares were triggered without coronal mass ejections. During the evolution of the active region, the flare emission seen in the H$\alpha$ and ultraviolet wavebands changed from a circular shape (plus an adjacent conjugated ribbon and a remote ribbon) to three relatively straight and parallel ribbons. Based on a series of reconstructed nonlinear force-free fields, we find sheared or twisted magnetic field lines and a large-scale quasi-separatrix layer (QSL) associated with 3D null points in a quadrupolar magnetic field. These features always existed and constantly evolved during the two days. The twist of the flux rope was gradually accumulated that eventually led to its instability. Around the flux rope, there were some topological structures, including a bald patch, a hyperbolic flux tube and a torus QSL. We discuss how the particular magnetic structure and its evolution produce the flare emission. In particular, the bifurcation of the flux rope can explain the transition of the flares from circular to parallel ribbons. We propose a two-stage evolution of the magnetic structure and its associated flares. In the first stage, sheared arcades under the dome-like large-scale QSL were gradually transformed into a flux rope through magnetic reconnection, which produced the circular ribbon flare. In the second stage, the flux rope bifurcated to form the three relatively straight and parallel flare ribbons. 

\end{abstract}
\keywords{Sun: flares --- Sun: UV radiation --- Sun: magnetic field --- Sun: magnetic topology}
\section{Introduction}

Solar flares are one of the most complex dynamic processes in the solar atmosphere, which release a large amount of energy in a wide wavelength range (e.g., \citealt{2011LRSP....8....6S}). They are usually associated with coronal mass ejections (CMEs) ejecting a large amount of magnetized plasmas. The magnetized plasmas along with energetic particles from explosive events propagate through the interplanetary space. They may impact the Earth, resulting in serious space weather conditions (e.g., \citealt{1991JGR....96.7831G,2011LRSP....8....1C,2017ScChE..60.1383C}). Therefore, it is important to study the origin of solar flares. In two dimensions, the formation and evolution of a flare is usually described by a standard model known as the CSHKP model (\citealt{1964NASSP..50..451C,1968IAUS...35..471S,1974SoPh...34..323H,1976SoPh...50...85K}). It states that a magnetic flux rope rises due to its instability (or loss of equilibrium), stretches the field lines above, and at the same time, leaves a current sheet behind where magnetic reconnection occurs. The rising flux rope could drive a CME, while the reconnection forms a series of post-flare loops that stride over the magnetic neutral line. Energetic particles can be accelerated during magnetic reconnection and then spiral down along the magnetic field lines to impact the lower atmosphere, where the flare ribbons are formed.\par

Although the standard 2D flare model is quite successful in interpreting some observations, flares indeed occur in three dimensions and show many features that cannot be described by the 2D model (\citealt{2012A&A...543A.110A}). Typical three-dimensional (3D) features include S-shaped coronal sigmoids \citep[e.g.,][]{2010ApJ...708..314A}, twisted flux ropes \citep[e.g.,][]{2009ApJ...693L..27C}, sheared and J-shaped flare ribbons \citep[e.g.,][]{2016ApJS..225...16C}, and circular flare ribbons \citep[e.g.,][]{2009ApJ...700..559M,2012ApJ...757..149S,2013ApJ...778..139S,2014SoPh..289.2041M}. Therefore, extensions of the 2D model, or quasi-3D models were proposed in the past two decades \citep[e.g.,][]{1995ApJ...451L..83S,2006A&A...453.1111T}. However, they still cannot explain all the new features revealed by the new observations with increasing resolutions. Recently, \citet{2010ApJ...708..314A} revealed the whole process of a flare, in which the formation and eruption of a 3D flux rope plays a key role in the flare process. First, bald-patch (BP) reconnection, driven by flux cancellation, transforms the sheared arcades into a flux rope. Second, due to the slip-running reconnection in the sheared coronal magnetic fields, the flux rope keeps rising and the BP separatrix is gradually transformed into a quasi-separatrix layer. Finally, when the flux rope reaches the critical point of the torus instability, it rises rapidly and triggers the eruption of the CME. 

In the standard flare model, the stability of the magnetic flux rope determines whether the flare is triggered or not. The magnetic flux rope is a very significant structure in the corona that is defined as helical magnetic field lines wrapping around the central axis with more than one turn. Magnetic flux ropes play important roles in triggering solar flares, driving CMEs, and structuring filaments. Over the past three decades, many authors have contributed to the study of magnetic flux ropes in various aspects including theory (e.g., see review by \citealt{2000JGR...10523153F}), observations \citep[e.g.,][]{2011ApJ...732L..25C,2012NatCo...3E.747Z}, and numerical simulations \citep[e.g.,][]{2000ApJ...529L..49A,2003ApJ...585.1073A,2003ApJ...595.1231A,2006ApJ...645..742I,2010ApJ...708..314A}. In particular, the scientific questions that have been explored include how the magnetic flux rope forms and evolves (\citealt{2010ApJ...708..314A,2012NatCo...3E.747Z}), whether a magnetic flux rope exists before a CME (\citealt{2015ApJ...809...34C}), how the magnetic flux rope becomes unstable (\citealt{2003A&A...406.1043T,2005ApJ...630L..97T,2016ApJ...832..106H}), how the magnetic flux rope breaks out (\citealt{1999ApJ...510..485A}), whether the magnetic flux rope is a necessary condition to produce a CME (\citealt{2015ApJ...815...72O}), and whether the magnetic flux rope is a supporting structure of a filament or not \citep{2010ApJ...714..343G,2011IAUS..273..328G}.\par

Furthermore, \citet{2012A&A...543A.110A} extended the standard flare model to three dimensions, especially on the nature of the strong-to-weak shear transition in post-flare loops by observations and 3D MHD simulations. They provided the pre-eruptive and the post-eruptive distributions of the shear quantitatively. \citet{2013A&A...555A..77J} further revealed some properties of 3D reconnection by MHD simulations, particularly the slip-running reconnection in quasi-separatrix layers, in eruptive flares. The standard model in three dimensions provides some predictions that have been confirmed by subsequent observations, for instance, a good correspondence between the photospheric footprints of the coronal electric current layer and the flare ribbons \citep{2014ApJ...788...60J}, and the apparent slip-running reconnection revealed by the motions of flare loops and S-shaped erupting loops in the same direction \citep{2014ApJ...784..144D}. Moreover, the motions of flare loops and S-shaped erupting loops in opposite directions along flare ribbons are also seen as an apparent slip-running reconnection \citep{2016ApJ...823...41D}.\par

The standard flare model in three dimensions does not contain null points. Therefore, it cannot explain a special type of flares, called circular ribbon flares, which are usually associated with the spine-fan structure and are widely studied recently \citep[e.g.,][]{2012ApJ...760..101W,2013ApJ...778..139S,2015ApJ...806..171Y,2017A&A...604A..76M}. Generally speaking, the morphology of a circular ribbon flare comprises a fully closed circular or quasi-circular ribbon, an elongated inner ribbon, and a remote ribbon. Photospheric line-of-sight magnetic field of the circular ribbon flares generally displays a small parasitic polarity being embedded in a main polarity of the opposite sign in an active region. In addition, circular ribbon flares have a special magnetic topology including a 3D null point, whose structure can be analyzed by a Taylor expansion and the Jacobian matrix in the vicinity of it (\citealt{1996PhPl....3..759P}). The Jacobian matrix has three eigenvectors corresponding to three eigenvalues, two of which have the same sign determining a fan separatrix, and the third one with the opposite sign defines a spine structure. The circular ribbon, inner ribbon, and remote ribbon are located at the footprints of the fan separatrix, inner spine below the fan separatrix, and the outer spine, respectively.\par

Magnetic reconnection can occur not only in the above configuration to produce circular ribbon flares \citep[e.g.,][]{2002JGRA..107.1164T,2017ScChE..60.1408G}, but also in other magnetic topological structures, such as bald patches (e.g., \citealt{1993A&A...276..564T,1998SoPh..183..369A,1999ASPC..184...65D}), separators \citep[e.g.,][]{2003SPD....34.0101L,2010JGRA..115.2102P,2010ApJ...725L.214P}, and, more commonly, quasi-separatrix layers (QSLs; e.g., \citealt{1995JGR...10023443P,2005A&A...444..961A}). QSLs are defined as places with strong gradients of magnetic connectivity \citep[e.g.,][]{1996A&A...308..643D,1996JGR...101.7631D,1997SoPh..174..229M,2002JGRA..107.1164T,2006AdSpR..37.1269D}. We usually characterize the gradient of magnetic connectivity through a topological parameter called the squashing factor $Q$ \citep{2002JGRA..107.1164T}. QSLs are places with a high value of $Q$, which is much larger than 2. Usually, QSLs appear as narrow layers with finite thickness. When magnetic reconnection occurs, the energetic particles along the field lines impact the footprints of QSLs, which intersect with the chromosphere, to brighten up flare ribbons (\citealt{1997A&A...325..305D,1997SoPh..174..229M}). These results demonstrate that we can analyze the evolution of flare ribbons to study the field line's linkage although reconnection mostly occurs in the corona. Furthermore, it is worth noting an essential structure called the hyperbolic flux tube (HFT; \citealt{2002JGRA..107.1164T}), which is located at the core of QSLs. The HFT structure is a generalization of a magnetic separator. It situates at the center of a flare region and corresponds to an X-type point in two dimensions, which could release a large amount of energy. Therefore, we can locate the possible position of magnetic reconnection effectively by these topological structures.\par

In some events, magnetic reconnection occurs in complex topologies that contain magnetic flux rope and various topological structures simultaneously. \citet{2013ApJ...778..139S} found that a twisted flux rope exists below the fan structure in a circular-ribbon flare, yet the flare ribbons corresponding to the flux rope do not show a separation due to its small size. \citet{2015ApJ...812...50J} found, however, a continuous separation between two parallel ribbons in a similar flare. In this event, a compact ribbon, two parallel ribbons, and a quasi-circular ribbon appear successively. They confirmed the existence of a flux rope and a null point. In many cases, multiple flare ribbons with different shapes are produced by complex magnetic topologies, rather than a simple structure such as a sheared two-ribbon flare corresponding to a magnetic flux rope or shear arcades and a circular ribbon flare associated with a null point. It should be noted that magnetic flux rope may not be simply shown as two foot points in the photosphere as we normally observe. A flux rope possibly connects multiple poles, or, to say, the flux rope might be bifurcated toward one or two ends. This is uncommon but actually appears in some observations. In such cases, it is impossible to describe the magnetic structure in two dimensions, while a 3D model is needed. On the other hand, we also need to study the interaction between the null points and the flux rope in order to understand how the flare is initiated and evolves.\par

In this paper, we make a detailed analysis of an active region containing a magnetic flux rope and null points that produced 6 M- and C-class flares from 2015 January 29 to 30. We construct a series of 3D magnetic field using a nonlinear force-free extrapolation and compute the topological structure in terms of the squashing factor $Q$. We then explore the formation of the magnetic flux rope and interpret the H$\alpha$ and extreme ultraviolet emission properties of the flares. The paper is organized as follows. In section 2, we describe the observations and modeling method. In Section 3, we present the results of the magnetic field and associated flare ribbons, and analyze the magnetic topology. We summarize major findings and give our discussions in Section 4.\par

\section{Observations and modeling}

\subsection{Overview of the active region}

The active region NOAA 12268 appeared on 2015 January 21 on the east limb of the Sun. At first, it was a simple active region including a negative polarity on the east and a positive polarity on the west. With the cancellation and emergence of the magnetic field, it gradually became a mosaic structure. A parasitic positive polarity appeared that was partially enclosed by the negative polarity on the east. Then, the active region produced the first flare of M1.1 class on January 26. Subsequently, the parasitic positive polarity continued to expand, and the biggest flare of M2.1 class occurred on January 29. Later, two flares of M2.0 and M1.7 class erupted on January 30. Finally, the flare activities decreased and ended at the west limb on February 4. Thus, this region was very active and flare-productive during January 29 and 30 (as shown in Figure 1). Since the active region was located near the central meridian of the Sun, the observed vector magnetic field on the photospheric boundary is accurate enough for further analysis. Here, we analyze the active region for 36 hours from 00:00 UT on January 29 to 12:00 UT on January 30. During this time, six flares (F1--F6) occurred in the active region (as listed in Table 1), which are all above C1.0 class as recorded by the soft-X ray instrument on board the \textit{Geostationary Operational Environmental Satellite} (\textit{GOES}). The evolution of the six flares was monitored by the ground-based solar telescope, Optical and Near Infrared Solar Eruption Tracer (ONSET; \citealt{2013RAA....13.1509F}), and Global Oscillation Network Group (GONG; \citealt{2011SPD....42.1745H}) in H$\alpha$. The six flares were also observed by the Atmospheric Imaging Assembly (AIA; \citealt{2012SoPh..275...17L}) on board the \textit{Solar Dynamics Observatory} (\textit{SDO}; \citealt{2012SoPh..275....3P}) with a high temporal cadence (12 s) and spatial resolution ($\sim 0.6\arcsec$ per pixel). Moreover, we carefully check the data observed by the Large Angle and Spectrometric Coronagraph (LASCO) C2 aboard the \textit{Solar Heliospheric Observatory} (\textit{SOHO}; \citealt{1995SoPh..162..357B}) and by AIA. We do not find any CME associated with the flares. Thus, the flares in this active region are regarded as confined ones and have attracted a lot of attention \citep[e.g.,][]{2017ApJ...847..124H}.\par

\subsection{Flare emission in H$\alpha$ and AIA multi-wavelengths}

During the time interval of 36 hours from January 29, the active region was very productive in flares. From the \textit{GOES} 1--8 \AA\ soft X-ray emission (Figure 1), we can see that flares larger than C1.0 class (including 3 M-class flares) occurred every six hours on average. Flare ribbons were observed in the H$\alpha$ and ultraviolet wavelengths corresponding to the chromospheric foot points of flare loops that have been formed by reconnection. These observations thus enable us to trace the morphological evolution of flare ribbons. In addition, the extreme ultraviolet images at 171 and 94 \AA\ from AIA provide another perspective of the active region. The emission of AIA 171 \AA \ originates from the upper transition region and the quiet corona, which contains the Fe \textsc{ix} line emission with a characteristic temperature of $6.3\times 10^{5}$ K. It usually reveals the low-temperature coronal structures. Different from 171 \AA, the 94 \AA\ waveband has a higher characteristic temperature of $6.3\times 10^{6}$ K. It includes the Fe \textsc{xviii} line emission from the flaring corona and is generally used to observe coronal structures with high temperatures. 

Figure 2a--2f and Figure 3a--3f show the AIA 1600 \AA\ and H$\alpha$ images of the six flares (F1--F6) at their peak time during the period from 00:00 UT on January 29 to 12:00 UT on January 30. All flare ribbons are labeled as R1--R6 in the AIA 1600 \AA\ images for each flare. Figure 3g--3l shows the six composite snapshots at 171 and 94 \AA\ corresponding to the \textit{GOES} peak flux times for the six flares. We divide the flare region into three parts: the east part (region E), the middle part (region M) and the west part (region W). The main characteristics of the six flares are summarized as follows.

\begin{enumerate}
\item The C8.2 class flare (Figures 2a, 3a, and 3g):
In the AIA 1600 \AA\ and H$\alpha$ images, two ribbons (R1 and R2) on each side of the polarity inversion line (PIL) appeared together at 05:15 UT. The \textit{GOES} flux reached its maximum at 05:23 UT. Subsequently, the two ribbons began to separate and ribbon R1 (the southern one) got elongated. Finally, the morphology of ribbon R1 changed to a semi-circular shape, and the flare looked like a circular ribbon flare. Ribbon R2 (the northern one) is the adjacent conjugated ribbon. With the evolution of the flare, the foot points of flare loops separated gradually from the PIL. In the AIA 171 \AA\ images, we find a lot of plasma flowing from regions E to M and some faint plasma flowing from regions M to E before the flare onset. Then, after the flare occurred, region M released a high temperature emission at AIA 94 \AA\ rapidly, which extended to both ends. Some plasma appeared flowing to region W simultaneously with the brightening of region M. At the same time, ribbon R3, a remote ribbon, was formed. Although the global picture looks like a circular ribbon flare, the evolution process of ribbons R1 and R2 can also be described by a strong-to-weak shear transition of flare loops predicted in the 3D flare model (\citealt{2012A&A...543A.110A}). 

\item The M2.1 class flare (Figures 2b, 3b, and 3h):
In this flare F2, the morphology of the flare ribbons was complex owing to multiple ribbons (R1, R2, R3, and R4). The emission characteristics at extreme ultraviolet wavebands were also different from that of flare F1. Some observations have already been described in \citet{2017ApJ...847..124H}. In contrast to flare F1, this flare has a much clearer remote ribbon (R4) in region W, and ribbon R2 in region M was not elongated. However, ribbon R1 appeared in region E. As seen in AIA 171 \AA, before the flare occurred, there was some plasma moving from regions M to E. Then, in the rising phase, region M released some emissions which extended to both ends in both AIA 94 \AA\ and 171 \AA. Moreover, we can see that some coronal loops were opened obviously and more loops were connected to regions M and W in the maximum phase. Meanwhile, the plasma flowed to region W with a wider brightening region than in flare F1. Region E showed a similar phenomenon to region W instead of a successive brightening caused by region M. \citet{2017ApJ...847..124H} interpreted such a phenomenon as magnetic reconnection in region M and plasma flow dissipation in regions E and W. 

\item The C6.4 class flare (Figures 2c, 3c, and 3i):
For this flare F3, the morphological evolution of the flare ribbons was similar to flare F1. Two ribbons (R2 and R3) also separated from each other during the flare evolution and the shape of the flare tended to be half-circular with the appearance of ribbon R1 in the end. Subsequently, a remote ribbon R4 was formed. However, there were still some differences. On the one hand, in H$\alpha$ image, ribbons R1 and R2 overlapped each other to form a ribbon that was much wider than ribbon R1 in flare F1. On the other hand, in the AIA multi-wavelength images, the low coronal loop connecting regions M and W had a structure that deviated from a semi-circular shape. Furthermore, high temperature emission at 94 \AA\ appeared not only in region E, but also in region W. 

\item The M2.0 class flare (Figures 2d, 3d, and 3j):
The morphology of this flare F4 was more complicated than the previous flares with four ribbons (R2, R3, R4, and R5) appearing in region M. After the occurrence of the flare, ribbon R2 in region M extended continuously and ribbon R1 was formed in region E. Ribbons R1 and R2 became a semi-circular shape. Meanwhile, a remote ribbon (R6) appeared with a narrow feature. The emission at 94 \AA\ had become visible before the flare occurred. The plasma flowed from the foot point in region M to the two foot points in regions E and W after the flare onset. Then, during the maximum phase, the multiple regions M, E, and W were brightened up at the same time. The flare in region M resembled an inverse S-shaped structure, implying that a flux rope came into existence. In addition, the lower coronal loops connecting regions M and W displayed extremely narrow structures that were different from the former three flares apparently. As a whole, the flare presented a complex structure including six ribbons in total. 

\item The C2.3 class flare (Figures 2e, 3e, and 3k):
In this flare F5, owing to its weakest strength among the six flares, the brightening of the flare ribbons was not obvious. However, great changes of the ribbon morphology had taken place compared to the aforementioned flares. On the one hand, the flare has three relatively straight and parallel ribbons instead of a circular ribbon (plus an adjacent conjugated ribbon and a remote ribbon). The sizes of each ribbon were closer to each other than the previous flares. We name such a flare as a parallel-ribbon flare hereafter. On the other hand, the distance between ribbons R2 and R3 became smaller than the aforementioned flares. Moreover, both ribbons R2 and R3 evolved to an elongated shape with time. In nor AIA extreme ultraviolet wavebands did the original circular characteristics exist. During the maximum phase, the inverse S-shaped structure in region M became more obvious than in previous cases. In the late phase, the inverse S-shaped structure connected regions M and W with a large amount of 94 \AA\ emissions. The whole structure was transformed into a flare with sheared ribbons. 

\item The M1.7 class flare (Figures 2f, 3f, and 3l):
This last flare F6 occurred half an hour after flare F5. It shows the apparent characteristics of a three-parallel-ribbon flare. The size of the remote ribbon (R3) was larger than that of the above flares. The morphology of the flare was completely different from that of the circular-ribbon flare. The flare was also different from previous three-ribbon flares in morphology (\citealt{2014ApJ...781L..23W,2014SoPh..289.2041M,2017ApJ...840...84S,2017ApJ...838..134B}). In our event, although all the three ribbons were near the PIL, they had different sizes and were only roughly parallel to each other. As seen in AIA 94 \AA, the inverse S-shaped structure in region M was brightening up instantaneously and released an intense emission during the maximum phase. At the same time, the emission at AIA 171 \AA\ was enhanced to show strong sheared flare loops. Then, region W released some emissions at AIA extreme ultraviolet wavebands, when ribbon R3 was formed.\par

\end{enumerate}

\subsection{Overview of photospheric vector magnetic field and preprocessing}

The active region was also well observed by the Helioseismic and Magnetic Imager (HMI; \citealt{2012SoPh..275..207S}) on board \textit{SDO}. \textit{SDO}/HMI observes the four Stokes profiles (\emph{I}, \emph{Q}, \emph{U}, and \emph{V}) of the Fe \textsc{I} 6173 \AA\ line. The photospheric vector magnetic field is derived by fitting the observed Stokes profiles using the Very Fast Inversion of the Stokes Vector (VFISV; \citealt{2011SoPh..273..267B}), which is provided by the HMI pipeline (\citealt{2014SoPh..289.3483H}) operated in the Joint Science Operations Center\footnote[1]{\url{http://jsoc.stanford.edu/ajax/exportdata.html}}. The data for study cover the full solar disk with a temporal cadence of 720 s and a spatial resolution of $1\arcsec$ with a plate scale of $0.5\arcsec$ per pixel \citep{2012SoPh..275..229S}. To analyze the vector magnetic fields, we make two more steps in addition to the processing of the HMI pipeline. First, we need to remove the 180$^\circ$ ambiguity of the transverse field following the minimum energy method \citep{1994SoPh..155..235M,2006SoPh..237..267M,2009SoPh..260...83L}. Second, we should correct the projection effect using the method proposed by \citet{1990SoPh..126...21G} since the active region is not located at the center of solar disk. Here, we coalign each frame of magnetogram to a $350\arcsec \times 290\arcsec$ region at 00:00 UT on January 29. As shown in Figure 3m--3r, we can see the vector magnetic field varying with time from 05:24 UT on January 29 to 05:36 UT on January 30. The overall structure of the active region was roughly maintained during this period. In the east, there was a positive polarity surrounded by a negative polarity, while a large positive polarity appeared in the west. With time elapsing, on the one hand, the parasitic positive polarity in the east was emerging continuously. In particular, the positive and negative polarities showed a sheared structure and were cancelled strongly with each other near the PIL before January 30. From January 30, the positive and negative polarities were inlaid with each other, thus accumulating a large amount of non-potential energy. On the other hand, a positive polarity in the west was also emerging continuously with time.\par

However, it is difficult to infer the connectivity of the positive and negative polarities only based on the photospheric vector magnetic field. Direct measurement of the coronal magnetic field is still immature at present. Considering that the plasma $\beta$ is well below unity, or the magnetic force dominates over the gas pressure, it is reasonable to assume a force-free state of the coronal field. Based on this fact, we can make use of the force-free field approximation to reconstruct the coronal magnetic field using the photospheric vector magnetic field as the bottom boundary.\par

In addition to the processing of the vector magnetic field on the photosphere as mentioned above, we need to address another problem. Due to the force-free field approximation, the vector magnetic field on the photopshere used for extrapolation should be force-free and torque-free. However, the vector magnetic field observed usually does not meet the above conditions. Thus, we apply the method proposed by \citet{2006SoPh..233..215W} to the observed vector magnetic field to remove the residual force and torque on the bottom boundary.\par

\subsection{Nonlinear force-free field modeling}

We use a nonlinear force-free field (NLFFF) model to reconstruct the coronal magnetic field. Here, we adopt the magneto-frictional method in the Message Passing Interface Adaptive Mesh Refinement Versatile Advection Code (MPI-AMRVAC; \citealt{2003CoPhC.153..317K,2012JCoPh.231..718K,2014ApJS..214....4P,2018ApJS..234...30X}). \citet{2016ApJ...828...82G} added the magneto-frictional module into MPI-AMRVAC and tested it with analytic solutions in both Cartesian and spherical coordinates. \citet{2016ApJ...828...83G} have successfully applied the method to observations in both Cartesian and spherical coordinates using the vector magnetic field observed by \textit{SDO}/HMI. This method only considers force-free field approximation with dissipation expressed as a friction form in the momentum equation. It solves the induction equation to find a state of magnetohydrostatic equilibrium simultaneously. In practice, we relax an initial non-force-free field, which is a combination of the extrapolated potential field and the observed photospheric field, to a force-free state. The potential field is computed from the normal component of the photospheric vector magnetic field. We use the preprocessed photospheric vector magnetic field as the inner bottom ghost layer boundary and adopt the zero-gradient extrapolation for the top, the outer bottom ghost layer and the other four sides in the Cartesian coordinate. Our NLFFF computational box is resolved into $ 240\times160\times160 $ grid points with a uniform distribution. Moreover, to evaluate the performance of the NLFFF model, two metrics proposed by \citet{2000ApJ...540.1150W} help us to check whether the model reaches the force-free state or not. One is the divergence-freeness metric and the other one is the force-freeness metric. In our extrapolation, the average of divergence-freeness metric is $\sim 1.5\times 10^{-4}$, and the average of force-freeness metric is $\sim 0.28$. The conditions of divergence-freeness and force-freeness are acceptable in applications of the method to observations. The two metrics in our results are similar to previous studies \citep[e.g.,][]{2008ApJ...675.1637S,2015ApJ...806..171Y,2016ApJ...828...83G}.\par

Although some metrics of the results are similar to the previous studies, the magneto-frictional method still has its inherent problems. In fact, some of these problems are common to other extrapolation methods. \citet{2009ApJ...696.1780D} compared four types of extrapolated methods. In their test event (AR10953), the finite-element Grad-Rubin method by AM1$^{-}$ not only has the least residual Lorentz forces and the least average divergence, but also has largest non-potential energy. In addition, magnetic field lines obtained by AM1$^{-}$ and the models of Wh$^{-}$, Am2$^{-}$ and Can$^{-}$ look like more aligned with the XRT loops than other models. However, through quantitatively comparing the NLFFF magnetic field lines and the \textit{STEREO} loops, it was found that none of the NLFFF models improves the angle metric in contrast to the potential field model. Then, \citet{2009ApJ...696.1780D} discussed various reasons, including the imperfect bottom boundary after preprocessing and the incomplete boundary information. Moreover, there is a problem that most of NLFFF algorithms do not converge to the force-free and divergence-free state perfectly. This problem also exists in the magneto-frictional method being used here. Although it is difficult to establish a coronal magnetic field that is perfectly consistent with observations, some indicators can be satisfied \citep{2009ApJ...696.1780D,2015ApJ...811..107D}, to say the force-freeness, divergence-freeness can reach the minimum if one adopts a bottom boundary as good as possible after doing an appropriate preprocessing. Meanwhile, the final reconstructed magnetic field can almost match the coronal image. In this case, we can still establish magnetic field models for effective estimation of physical quantities.\par

Based on the vector magnetogram at 00:36 UT on January 30, we test a part of free parameters in the magneto-frictional method that are related to the relaxation process to reconstruct NLFFF. We also discuss the effects on reconstructing NLFFF of changing a series of free parameters for four aspects. First, the linear proportional coefficients $c_c$ and $c_y$, which control the magnitude of the magneto-frictional velocity in Equation (6) in \citet{2016ApJ...828...82G}, have no significant influence on the reconstructed magnetic field, unless the coefficients are too large, that is, the frictional velocity cannot be guaranteed less than or equal to the largest possible information propagation velocity. Secondly, the coefficient $1/|B|^{2}$ in the equation of the magneto-frictional velocity has more influence on the convergence speed of the weak-field regions. If we change this coefficient to, for example, $1/|B|$, we need more time to make the weak-field regions converge. Thirdly, the function $f_w$(x) ensures the frictional velocity in the four side boundaries and the top boundary to reach a lower value, which only affects the magnetic field in the boundary. Since the field of view we selected is large enough that contains magnetic connectivities as many as possible, for the center of the active region, the function also has little influence on the reconstructed magnetic field. Finally, the boundary conditions with different spatial resolutions can affect the NLFFF modeling results \citep{2015ApJ...811..107D}. During the process of reconstructing the magnetic field, we use the bottom magnetogram with a resolution as high as possible within our computation capabilities. Meanwhile, we ensure that the final magnetic field is close to a force-free state.\par

In addition, during the process of reconstructing the magnetic field, how to choose the stopping point is a problem that must be confronted. The stopping point is chosen as the iteration step at which the force-free metric is minimized. There is also a problem that the magneto-frictional relaxation does not converge perfectly to the force-free and divergence-free state overall, and even diverges at some specific weak field regions. We have tried to solve these problems by our method. First of all, our choice of the stopping point is not arbitrary. We have run the code for a long time until the two metrics of force-freeness and divergence-freeness begin to diverge. Then we choose the position where the force-free metric reaches the minimum value as the stopping point. At the same time, we compare the magnetic field that runs for a long time with the magnetic field at the chosen stopping point for each event. We find that both magnetic field structures are similar in the strong field regions. The divergence exists in the boundary and weak field regions for both cases, although the case stopping at the minimum metric has weaker divergence than the case with a long time iteration. These regions with divergence are located at the grids with weak fields fluctuating between positive and negative. We cannot deal with this problem due to the imperfect extrapolation method. Fortunately, these structures only appear in a few weak field regions, and do not affect the structure of the overall magnetic field.

For the boundary data, we choose a region that is as large as possible to cover all magnetic connectivities and include the strong field region with the flux balance of the bottom boundary being guaranteed. Thus, the magnetic field lines in our region of interest are self-closed on the bottom boundary. It ensures the reliability of the reconstructed magnetic field in the center region of the six events. Then, the bottom boundary removes net forces and torques, which is much closer to a force-free state after preprocessing. Nevertheless, there is a number of free parameters in the preprocessing, namely $\mu_1$, $\mu_2$, $\mu_3$, and $\mu_4$, which are defined by \citet{2006SoPh..233..215W}. The four free parameters are the weights for the force-free condition, the torque-free condition, the deviation to the observation, and the smoothness, respectively. Although \citet{2006SoPh..233..215W} provided optimal parameters for each of them, i.e. $\mu_1$ = 0.1, $\mu_2$ = 0.1, $\mu_3$ = 0.0001, and $\mu_4$ = 0.001. These parameters do not necessarily apply to all observations. Therefore, we test two of the four parameters in the preprocessing code for the vector magnetogram at 00:36 UT on January 30, and then we examine how the magnetogram preprocessed by different parameters affects the extrapolation results. First, we fix $\mu_1$ and $\mu_2$, and change $\mu_3$ and $\mu_4$, respectively. We find that the bifurcated magnetic flux rope can be reconstructed within a certain range of the parameters. Only when $\mu_3$ or $\mu_4$ deviates too far from the original optimization value, does the structure of the flux rope change a lot. Secondly, we add artificial noises, whose root-mean-square is equivalent to the observed one, to the preprocessed magnetic field. We find that the morphology of the reconstructed magnetic field is almost identical to the original magnetic field without additional noisy structures.\par

\section{Results}

\subsection{Reconstruction of the coronal magnetic field}

With the aforementioned critical assessment of the NLFFF models, we reconstruct a series of 3D coronal magnetic field from 00:00 UT on January 29 to 12:00 UT on January 30 in the Cartesian coordinate. Figure 4 shows the AIA extreme ultraviolet images overlaid with the coronal magnetic field lines at the moment of the flux rope bifurcation. In general, a magnetic flux rope is a special structure in which all magnetic field lines wrap around the central axis with more than one turn. A simple magnetic flux rope is generated by a current channel that has only a poloidal component. The current channel has two foot points rooted in the photosphere. Many previous magnetic flux rope models are based on this idea. However, in some accidental situations, when magnetic reconnection occurs, the current system can be changed and the magnetic flux rope is no longer isolated. One possible result is the bifurcation of the magnetic flux rope after magnetic reconnection. For example, the bifurcated flux rope can have two current systems with different directions and one of the foot point can be divided into two.

In order to show the magnetic field lines at both high and low atmospheric layers, we present a full disk image and a zoomed-in image for a local area, respectively. The image at 171 \AA\ waveband is sensitive to a low temperature of quiet corona ($6.3\times 10^{5}$ K). Because of the frozen-in effect in the corona, we can roughly think that the coronal loops are shaped by the magnetic field lines. From the extrapolated field lines and the 171 \AA\ images, we can see that they are almost co-spatial (Figure 4a). The images at 94 \AA\ waveband usually reflect the high temperature of flaring plasma ($6.3\times 10^{6}$ K). A magnetic flux rope has usually a strong current along its axis and is thus more visible in the high-temperature wavebands. From Figure 4b, it is seen that the 171 \AA\ emission is not obviously strengthened at the position of the magnetic flux rope. However, the magnetic flux rope can be well traced by the 94 \AA\ emission in Figure 4c. It thus confirms that the magnetic flux rope corresponds to a high temperature structure. In some previous studies, the main body of a flux rope was thought to manifest as a hot channel structure visible in AIA high-temperature wavebands (\citealt{2012ApJ...761...62C,2014ApJ...789...93C}), no matter whether there is a filament or not. Our results are consistent with the previous finding.\par

We further analyze the magnetic field before each flare in a computational box with $240 \times 160 \times 160$ grids. For this purpose, we overlay the magnetic field lines on the images at the ultraviolet wavebands. As shown in Figure 5, there is a dome-like structure in the corona for each flare. The size and shape of this structure did not change too much during the period of 36 hours. Through a topological analysis, we find several null points in the dome-like structure. Moreover, we pay attention to the change of field lines in the low atmosphere. We propose a two-stage evolution of the magnetic structure.\par

In the first stage, sheared arcades were gradually transformed into a flux rope by magnetic reconnection. Flare F1 occurred at 05:15 UT on January 29. There were several sheared arcades being formed in region M (Figure 5a). The height of the magnetic arcades is about 21.5 Mm from the photosphere. In addition, the magnetic field lines extending from the western positive polarity did not connect with the sheared arcades in region M under the dome-like structures. At 11:32 UT, flare F2 occurred and the low magnetic arcades became more sheared than the former state (Figure 5b). In particular, some foot points of sheared arcades were very close to the foot points of the field lines extending from the western positive polarity. It implies that the magnetic arcades might have interacted with the field lines extending from the western positive polarity due to magnetic field emergence and cancellation in the photosphere. With the time going on, the low magnetic field has undergone a great change after eight hours and flare F3 occurred at 19:36 UT. The magnetic flux rope had been formed as seen from the reconstructed coronal magnetic field (Figure 5c). The height of the magnetic flux rope is nearly 20 Mm. At this moment, the twist of the magnetic flux rope was not high, and the curvature of the magnetic field lines wrapping around the central axis was not large. Meanwhile, the magnetic field lines extending from the western polarity still kept interacted with the sheared magnetic arcades. Some lower field lines no longer displayed a structure close to a semicircle like the previous two events. There were already some twists in those field lines.\par

In the second stage, a bifurcated magnetic flux rope was formed with the appearance of multiple flare ribbons, and the twist of the rope was gradually accumulated that finally led to its instability. Flare F4 occurred at 00:32 UT on January 30. The shape of the magnetic flux rope was totally different from the initial flux rope (Figure 5d). The twist of the magnetic flux rope was larger than before, and the curvature of the magnetic field lines wrapping around the central axis became larger, too. Furthermore, the magnetic field lines primarily linking regions E and M were connected to the western positive polarity through magnetic reconnection, forming a BP structure. The eastern and the western magnetic polarities that were initially unconnected became connected with each other completely. About four hours later (04:51 UT), the bifurcated flux rope continued to develop (Figure 5e). The magnetic field lines extending from the western positive polarity became a part of the magnetic flux rope, and they also provided some twist to the flux rope. The magnetic flux rope looked like a solenoid with a strong twist. However, this magnetic flux rope with more twist only produced a C2.3 class flare (F5), much smaller than previous flares (F1--F4). Only about half an hour later (05:29 UT), an M1.7 class flare (F6) exploded from the active region again. At this moment, the twist of the bifurcated flux rope decreased (Figure 5f). It implies that the flare was caused by the eruption of the magnetic flux rope. However, we did not find any CME in the LASCO coronagraph after the eruption of the magnetic flux rope. Two intriguing questions, namely how the magnetic flux rope was triggered and why it did not produce any CME, are beyond the scope of this work. We will make a further analysis on this problem in the follow-up work.\par

\subsection{Magnetic configurations associated with flare ribbons}

Flare ribbons correspond to the chromospheric foot points of flare loops that have formed through reconnection. We can explain the behavior of flares by the reconstructed 3D coronal magnetic field. However, there are two limitations with the available data. One is that the resolution of the vector magnetogram is lower than that of the AIA images. The other is that the magnetic field evolves dynamically during the occurrence of flares. Thus, the real magnetic field does not necessarily satisfy the force-free condition at that time. Here, we only use the reconstructed magnetic field based on the vector magnetograms before the flare occurred. Figure 5 shows the apparent motion of the flare ribbons at AIA 1600 \AA\ waveband. Flare ribbons changed from a circular shape (plus an adjacent conjugated ribbon and a remote ribbon) to six complex ribbons, and to three relatively straight and parallel ribbons in the end. We describe how the different shapes of flare ribbons are generated in different magnetic environments. Meanwhile, we compute the positions of magnetic null points by solving the equation $B_{i}(x, y, z) = 0$ (where $i = x, y, z$). The null points found in the domain of the six events are listed in Table 2. The magnetic configurations associated with ribbons of each flare are summarized below.

\begin{enumerate}
\item Flare F1 (Figures 5a and 6a):
For this flare F1, we find two null points (a and b) in our computational box. The circular shape could be explained by null points. But the dome-like structure was not closed completely. A more detailed topology analysis reveals that the dome-like structure was not a fan structure. In addition, observations reveal that the circular flare ribbon was composed of two main ribbons (R1 and R2) and a faint ribbon (R3). Ribbons R1 and R2, which were located at the foot points of a series of magnetic sheared arcades, had an obvious separation. Therefore, we think that the brightening of the circular ribbon was mostly produced by the magnetic reconnection of the sheared arcades. At the same time, null point reconnection has played a partial role in the extension of ribbon R1. Moreover, because of the weak spatial relationship between the sheared arcades and the field lines extending from the western polarity, the brightness of ribbon R3 was low in region W.\par 

\item Flare F2 (Figures 5b and 6b):
In this flare F2, its morphology displays four ribbons. The mechanism of the flare has been discussed by \citet{2017ApJ...847..124H}. They proposed that the flare ribbons in region M were caused by magnetic reconnection due to non-thermal particle acceleration. And flare ribbons at both sides of the flaring region were interpreted as plasma kinetic energy dissipation. We present an alternative explanation to this flare F2. In this flare, only one null point (about 3.3 Mm high) is found in the domain. We find that ribbon R1 in region E was connected with null point a and ribbon R4 in region W was linked with the foot points of the sheared arcades from the reconstructed magnetic field. Ribbons R1 and R2, which were located at the foot points of a series of magnetic sheared arcades, also had an obvious separation. Combined with the evolution of flare ribbons seen in AIA 1600 \AA, we find that ribbons R2 and R3 started to brighten at first, followed by ribbon R4, and finally by ribbon R1. Therefore, we consider that magnetic reconnection in the sheared arcades induced flare ribbons R2 and R3 in the first stage. Subsequently, magnetic reconnection occurred in null point a and the QSLs extending to regions E and W, causing flare ribbons R4 and R1 to brighten up one after the other.

\item Flare F3 (Figures 5c and 6c):
We find two null points (a and b) and a magnetic flux rope (green lines in Figure 5c) in flare F3. Although the flux rope had been formed, the behavior of the ribbons R2 and R3 caused by the flux rope is similar to that caused by the magnetic arcades. We also think that null point reconnection makes a little contribution to main ribbons R2 and R3, but plays a partial role in the extension of ribbon R1. In addition, due to the reconnection between the sheared arcades and the field lines extending from the western polarity, the remote ribbon (R4) brightened up subsequently.

\item Flare F4 (Figures 5d and 6d):
With the evolution of flare F4, its morphology tended to be more complex, consisting of both circular and sheared ribbons. To sum up, the flare comprised six ribbons. From the reconstructed magnetic field, we find that a good correspondence between the foot points of magnetic field lines and flare ribbons. We also find two null points (a and b) above ribbons R3 and R5. Combined with the separation of the flare ribbons and the bifurcated flux rope, we consider that the magnetic configurations went through the following changes. First of all, ribbons R2 and R4 were caused by the flux rope in region M, while the north ribbons (R3 and R5) were induced by null point reconnection. Almost at the same time, the eastern ribbon (R1) was formed by the impact of particles along the field lines which are related to null points and the flux rope. Similarly, we think that null point reconnection only contribute a small part. Owing to the bifurcation of the flux rope, the energetic particles could also propagate along the bifurcated field lines, and eventually impact the low atmosphere, causing the remote ribbon (R6). 

\item Flare F5 (Figures 5e and 6e):
The flare ended up with three relatively straight and parallel ribbons. There occurred a weak flare (F5) and a strong flare (F6). In flare F5, the circular features no longer existed and the ribbons looked faint. However, the reconstructed magnetic field displayed the bifurcated flux rope clearly, corresponding to the three ribbons (R1, R2, and R3). In addition, although we find three null points (a, b and c), we consider that null point reconnection contributed little to the brightening of flare ribbons, which was mainly caused by the bifurcated flux rope. As shown in Figure 4c, the bifurcated flux rope was also consistent with the high temperature structure at AIA 94 \AA\ waveband. It indicates that the dynamic evolution of the flux rope, its bifurcation, did occur in the coronal magnetic field. These observations cannot be described by the standard 2D or 3D flare models. They could possibly be explained by a multi-footpoint magnetic flux rope model. 

\item Flare F6 (Figures 5f and 6f):
This strong flare took place about half an hour later. The released energy was eight times larger than the former. Therefore, the features of the three-parallel-ribbon flare caused by the flux rope bifurcation were very obvious. We find four null points (a, b, c and d), which are considered to contribute a part to the extension of ribbon R2. We note some differences in flare ribbon evolutions between our flare and the three-ribbon flares reported in previous papers (\citealt{2014ApJ...781L..23W,2014SoPh..289.2041M,2017ApJ...840...84S,2017ApJ...838..134B}). In fact, this is due to a fundamental difference in the magnetic field. In our case, due to the different magnetic structure of the flare region, the three-parallel-ribbon flare could not be explained by the above scenarios. From the reconstructed magnetic field, we proposed that the three-parallel-ribbon flare is caused by a specific magnetic flux rope with three foot points. The three foot points could correspond to the three flare ribbons.
\end{enumerate}

\subsection{Magnetic reconnection in the flux rope formation and bifurcation}

Although we can observe the formation and bifurcation process of the magnetic flux rope from the reconstructed magnetic field, it is still essential to find the place where magnetic reconnection occurs, by analyzing the magnetic topology for the active region. We compute the squashing factor $Q$ by the method proposed in \citet{2012A&A...541A..78P} and obtain the QSLs for each flare. During the period of 36 hours, the QSLs did not change significantly as a whole except in the low atmosphere. Magnetic reconnection occurred in sheared magnetic arcades across the PIL in the low atmosphere. Therefore, we choose typical cases (flares F4 and F5) for illustration that contains all the topological structures at the flux rope bifurcation (as shown in Figure 7). The QSL distribution clearly reveals that the active region has a complicated topology corresponding to a large-scale QSL and a flux rope surrounded by a torus QSL. The large QSL is induced by the quadrupolar-like magnetic field. We can clearly see a positive polarity, surrounded by a negative polarity in region E, and a large positive polarity in region W. Owing to the existence of the magnetic flux rope, the semicircular-shaped negative polarity is separated into two parts. Thus, two negative polarities with two associated positive polarities constitute a quadrupole field.\par

We also find a torus QSL tightly around the bifurcated flux rope. Figure 7a, 7b, 7d, and 7e clearly shows the morphology of the torus QSL below the large-scale QSL from two different directions. The torus QSL is much lower and shows a high $Q$ value where the magnetic flux rope bifurcates. It implies that magnetic reconnection is prone to occur in the bifurcation position. Meanwhile, with a higher twist of the bifurcated flux rope, the torus QSL displays a higher $Q$ value. Moreover, we overlay the QSLs on the flare ribbons (Figure 6). We find that the QSLs match the flare ribbons very well in the six events. It is similar to the previous study in \citet{2015ApJ...810...96S}. It is known that QSLs provide possible acceleration locations for energetic particles that can impact in lower solar atmosphere to form the flare ribbons.\par

In addition to the overall topological structure, we also analyze the specific process of the magnetic flux rope formation and bifurcation. First, we study the role of magnetic reconnection in the formation of the magnetic flux rope. Figure 8a clearly displays that two green field lines in opposite directions form a red field line by BP reconnection at a position tangent to the photosphere. The QSL slices display that the $Q$ value around the flux rope is higher than in the surroundings (Figure 8b). Secondly, we explore the specific reconnection process leading to the flux rope bifurcation. As shown in Figure 8c, two green field lines that were originally disconnected in both sides were finally linked by the BP reconnection. Then, the western positive polarity was connected to the flux rope. We also find that the $Q$ value at the BP positions was higher than that in the background (see the QSL slice in Figure 8d). Therefore, magnetic reconnection is easy to occur at the bifurcation position.\par

In order to view the topological structure clearly after the flux rope bifurcation, we make a number of vertical cuts perpendicular to the flux rope axis. As shown in Figure 8e, the $Q$-map displays the QSL footprints in the photosphere. We find multiple topological structures in the QSLs (Figure 8f--8h). First, a highly structured HFT is situated above the flux rope. It can also be seen in Figure 7d. The HFT structure in the large QSL induced by the quadrupolar-like field is similar to the theoretical scenario in \citet{2002JGRA..107.1164T}. In addition to the presence of a HFT and null points in the higher atmosphere, we also find both a BP structure (Figure 8g) and a low HFT structure (Figure 8h) in the lower atmosphere. Both the low and high HFT structures are favorable places for magnetic reconnection.\par

Magnetic reconnection in the lower solar atmosphere is usually driven by magnetic flux emergence and cancellation. With the evolution of the magnetic fluxes, we find some evidence of the flux emergence and cancellation at the PIL. Overall, the positive flux and the absolute value of the negative flux in the middle of the active region kept increasing during the period of 36 hours. However, these fluxes turned to decrease during some flares, which provides clear evidence of flux cancellation. For instance, the absolute values of the negative flux decreased during flares F2 and F4, but increased during flare F5. This signature was not clear for flares F1, F3, and F6. We note that magnetic reconnection in flares F2 and F4 occurred in the BP, while that in flare F5 occurred in the HFT, which is consistent with the flux cancellation and flux emergence model (e.g., \citealt{2010ApJ...708..314A,2003PhPl...10.1971L,2008A&A...492L..35A,2009ApJ...691.1276A}).

\section{Summary and Discussion}

We investigate the formation and bifurcation of a magnetic flux rope responsible for a series of solar flares in a complicated active region during 00:00 UT on January 29 to 12:00 UT on January 30. Six solar flares were clearly observed by several instruments including ground-based ONSET and GONG, and space-borne \textit{SDO}/AIA and HMI. The magnetic flux rope exhibits a hot channel structure and a complex magnetic topology. To understand the characteristics of the six flares, we reconstruct a series of coronal magnetic fields with the nonlinear force-free field model and calculate the values of the squashing factor $Q$. The main findings of the paper are summarized as follows.\par

\begin{enumerate}
\item{Based on the flare emission seen in the H$\alpha$ and ultraviolet wavebands, we identify that the flare ribbons transformed from a circular shape (plus an adjacent conjugated ribbon and a remote ribbon) to multiple ribbons, and became three relatively straight and parallel ribbons eventually. It demonstrates that the field line's linkage underwent a tremendous change during the period of the six different flares.}

\item{The evolution of the magnetic structure and its associated flares went through two stages: (1) A magnetic flux rope under the dome-like large-scale QSL was formed by magnetic reconnection in sheared arcades driven by magnetic flux emergence and cancellation. The flux rope produced a circular ribbon flare. (2) The flux rope tended to be bifurcated, resulting in a multiple ribbon flare. The twist of the bifurcated flux rope was accumulated by reconnection that was accompanied with a parallel-ribbon flare with three ribbons in the end.}

\item{We analyze the complex magnetic topology associated with the bifurcated flux rope. We find a large QSL with 3D null points that stably existed during the period of 36 hours. It explains why the eastern flare ribbon (ribbon R1 in Figure 2a, 2c, and 2d) was long and circular. It can also interpret the appearance of several small flare ribbons (ribbons R2 and R4 in Figure 2d). With the bifurcation of the flux rope, a BP, a lower HFT and a torus QSL surrounding the flux rope also appeared. These structures are favorable positions for magnetic reconnection.}

\end{enumerate}\par

Based on the observed results, we propose a scenario for the magnetic flux rope formation and bifurcation as shown in Figure 9. First, with the flux cancellation and shearing motion of the photospheric vector magnetic field, the originally potential magnetic arcades become gradually sheared around the PIL. When the sheared arcades come into contact with each other, tether-cutting or BP reconnection is supposed to happen between them. Secondly, a magnetic flux rope is gradually formed by magnetic reconnection between the sheared arcades. At the mean time, non-potential magnetic energy is gradually accumulated. Thirdly, the magnetic flux rope becomes more twisted and interacts with the surrounding magnetic field through BP reconnection. At this moment, a bifurcated flux rope has been formed. It connects three magnetic polarities and continues to accumulate twists. Finally, the bifurcated flux rope with a high twist continues to go up. In our observations, we do not find CMEs after the eruption of the magnetic flux rope. We need a further analysis about how the magnetic flux rope is triggered but restricted.\par

Flare ribbons correspond to the chromospheric foot points of flare loops. Using H$\alpha$ and high-resolution ultraviolet observations, we identify that flare ribbons changed morphologies during the period of the six flares, first a circular shape, then multiple ribbons, and at last three ribbons in active region NOAA 12268. It suggests that the magnetic structure has also undergone a drastic change. At the beginning, a circular ribbon implies existence of a null point in the corona. Later, appearance of two ribbons suggests that magnetic reconnection occurs in the magnetic arcades or the flux rope. Moreover, coexistence of three or more ribbons illustrates that a variety of magnetic structures could interact with each other. We note that, although the circular ribbon appeared in some of the flares, the ribbon was not closed completely. It implies that the ribbon may not correspond to a fan structure. Based on the topology analysis, we find that the circular ribbon was linked with several spine field lines corresponding to a few null points. The magnetic topology near the null points was complicated due to their interaction with the flux rope. Below the null points, we can see that the flare ribbons (R2 and R4) brightened up with the fan structures (Figures 5d and 6d). Namely, circular ribbon in region E was not corresponding to fan structure. This result is different from previous studies (\citealt{2012ApJ...760..101W,2013ApJ...778..139S,2015ApJ...806..171Y}), where the circular ribbons correspond to the magnetic fan structure. In addition, these results provide a reasonable interpretation to the question why region E has an individual ribbon (R1) instead of continuous ribbon in flare F2 (Figures 2b and 5b). \par

We also note that the three sheared ribbons require a completely new model other than those in previous studies (\citealt{2014ApJ...781L..23W,2014SoPh..289.2041M,2017ApJ...840...84S,2017ApJ...838..134B}). \citet{2014ApJ...781L..23W} reported two three-ribbon flares, which occurred in succession and were located at the both sides of the flaring region. The three ribbons of the first flare had almost the same size and ran parallel to the PIL. The second flare appeared at the extended position of the first flare. They revealed that the three-ribbon flare was caused by reconnection along the coronal null line. \citet{2014SoPh..289.2041M} mentioned an asymmetrical three-ribbon flare including an irregular circular ribbon with a bright bar. They provided a proper interpretation to the three-ribbon flare. The flare was caused by magnetic reconnection at the separatrix that was formed by multiple asymmetric null points. \citet{2017ApJ...840...84S} revealed a thin three-ribbon flare, which corresponds to a small-scale magnetic arcade in the photosphere. \citet{2017ApJ...838..134B} showed another three-ribbon flare triggered by small magnetic disturbances with reversed shear. The three ribbons had irregular shapes and were partially parallel to each other. They interpreted it as magnetic reconnection by two groups of the magnetic structures. In these studies, the three ribbons were explained in accordance with their own observational characteristics. At present, there is no unified model to describe three-ribbon flares due to the complexity of the magnetic structures. Usually, the standard 2D model (CHSKP model) can only explain the two-ribbon flares. In the 3D flare model \citep{2012A&A...543A.110A,2013A&A...555A..77J,2014ApJ...788...60J,2016ApJ...823...41D} proposed in recent years, the key element is a magnetic flux rope that has two foot points. Although the magnetic flux rope is itself three dimensional, it can still explain existence of two flare ribbons, though they can be highly sheared in many cases. Therefore, a specific magnetic flux rope with bifurcated foot points provides a possible explanation to some flares with multiple ribbons as the one presented in this paper.\par

From AIA extreme ultraviolet observations, we also study the formation process of the twisted structure. We find the transformation from sheared magnetic arcades to the magnetic flux rope. In the earlier events (F1--F3), flare loops display a strong-to-weak shear transition. It is consistent with the 3D flare model proposed by \citet{2012A&A...543A.110A}. In the latter events (F4--F6), an inverse S-shaped structure with high temperatures, also known as a hot channel, is revealed in the extreme ultraviolet images. It is regarded as a manifestation of the magnetic flux rope.\par

There are still some problems to be noticed in the above results. First, we can get very small force-freeness and divergence-freeness metrics, but the magnetic field still does not converge very well in some positions. As discussed in Section 2.4, these positions of divergence are located in weak field regions or boundaries. Although these small regions of divergence do not affect the structure of the overall magnetic field, we have not solved this problem completely. We will continue to optimize the magneto-frictional algorithm in the future to solve the problem of the divergence in weak field regions. Secondly, the magnetic field is based on the quasi-static force-free field model. In fact, we do not know the real evolution of the magnetic field during the flare period that should be highly dynamic. With \textit{SDO}/HMI, high-cadence, high-resolution magnetograms can be acquired. We can use a time series of magnetograms as a boundary condition to drive the evolution of the coronal magnetic field. Such data-driven simulations will be much closer to the real situations. In conclusion, we discover a bifurcated magnetic flux rope in the solar atmosphere that has not been reported before and study its role in the production of a number of flares. We expect to perform data-driven simulations to explore more clearly how the flux rope is formed and the flares are produced.\par

\acknowledgements We are grateful to the referee for constructive comments that helped improve the paper. We thank the Optical and Near Infrared Solar Eruption Tracer (ONSET) team, \textit{Solar Dynamics Observatory} (\textit{SDO}) team and \textit{Solar Heliospheric Observatory} (\textit{SOHO}) team for providing the data for study. ONSET is a ground-based solar telescope of Nanjing University installed at the Fuxian Solar Observatory. \textit{SDO} is a mission of NASA's Living With a Star Program. \textit{SOHO} is a project of international cooperative effort between ESA and NASA. This work was supported by NSFC under grants 11373023, 11733003, 11773016, 11703012, 11533005, Jiangsu NSF under grants BK20170619, and the fundamental research funds for the central universities 020114380028. Magnetic field lines are visualized with ParaView (https://www.paraview.org).

\newpage
\begin{deluxetable}{ccccccc}
\centering
\tablewidth{0pt}
\tablecolumns{7}
\tablecaption{List of flares in active region NOAA 12268\tablenotemark{a} \label{t:1}}
\tablehead{
\colhead{Flare} & \colhead{SOL\tablenotemark{b}} & \colhead{Onset time [UT]} & \colhead{Peak time [UT]}   & \colhead{Class} & \colhead{CME} & \colhead{Location}}
\startdata
F1 & SOL2015-01-29T05:15:00L046C103    &       05:15  	& 05:23    & C8.2     & no   &S13W03 \\
F2 & SOL2015-01-29T11:32:00L042C102    &       11:32  	& 11:42    & M2.1     & no   &S12W09 \\
F3 & SOL2015-01-29T19:30:00L032C090    &       19:30  	& 19:44    & C6.4     & no   &S13W13 \\
F4 & SOL2015-01-30T00:32:00L029C090    &       00:32  	& 00:44    & M2.0     & no   &S13W16 \\
F5 & SOL2015-01-30T04:51:00L027C090    &       04:51  	& 04:55    & C2.3     & no   &S12W21 \\
F6 & SOL2015-01-30T05:29:00L027C090    &       05:29  	& 05:36    & M1.7     & no   &S12W19 \\
\enddata
\vspace{0.1in}
\tablenotemark{a}{All the flare parameters are extracted from the \textit{GOES} flare list and CMEs are checked by the observations of \textit{SOHO}/LASCO and \textit{SDO}/AIA.}\par
\vspace{0.05in}
\centerline{\tablenotemark{b}{The six flares are identified using solar observation target identification convention \citep{2010SoPh..263....1L}.}}
\end{deluxetable}
\newpage
\begin{figure} 
      \vspace{-0.0\textwidth}    
      \centerline{\hspace*{0.00\textwidth}
      \includegraphics[width=1.0\textwidth,clip=]{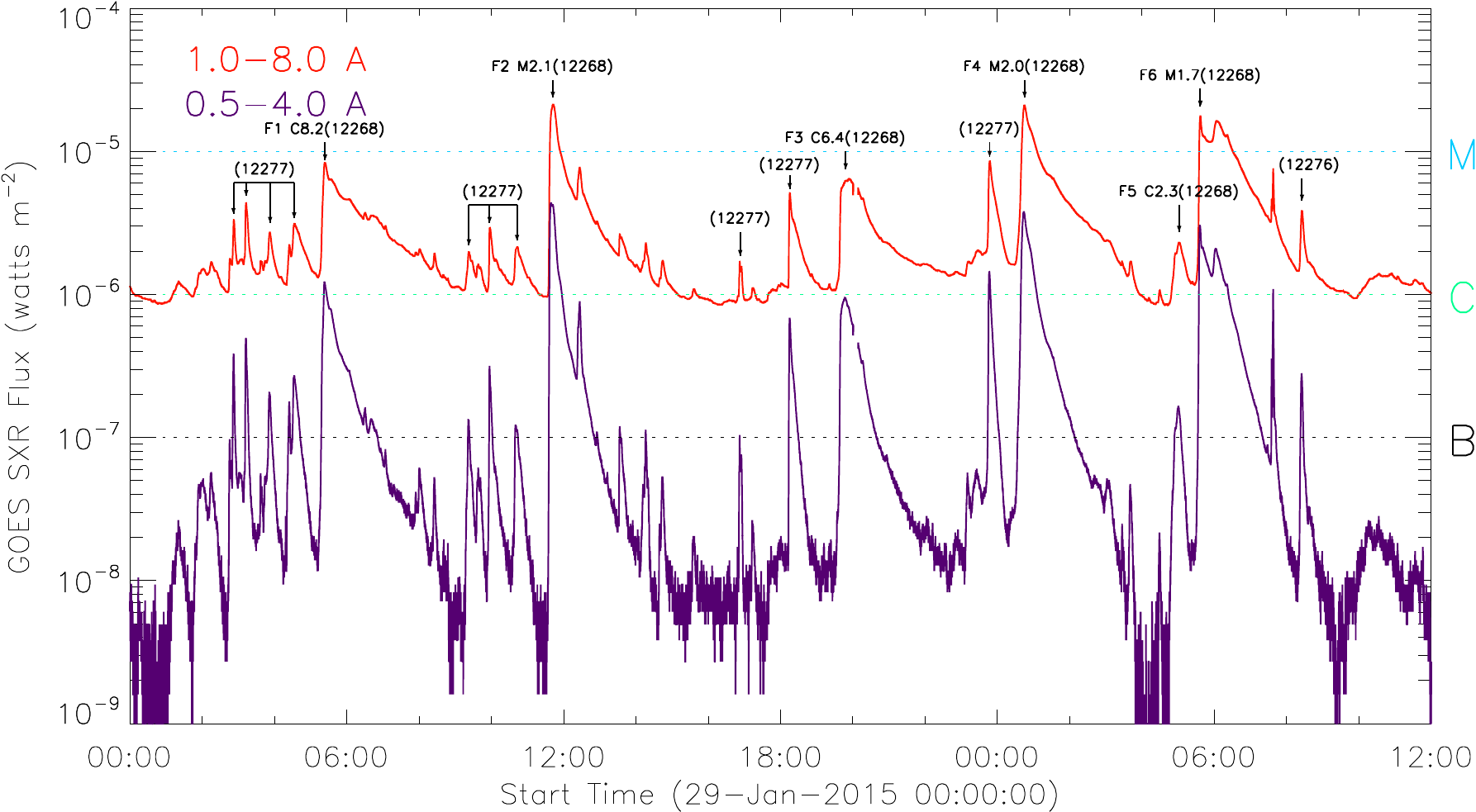}
      }
\caption{ \textbf{Soft X-ray flux observed by \textit{GOES} from 00:00 UT on 2015 January 29 to 12:00 UT on 2015 January 30}. Black arrows show flares greater than C1.0 that occurred in active regions (with corresponding numbers) during the period. The classes of the six flares in active region NOAA 12268 are marked. Red and purple curves represent the soft X-ray fluxes at 1.0--8.0 \AA\ and 0.5--4.0 \AA, respectively.}
\label{fig:1}
\end{figure}
\begin{figure} 
      \vspace{-0.0\textwidth}    
      \centerline{\hspace*{0.00\textwidth}
      \includegraphics[width=1.0\textwidth,clip=]{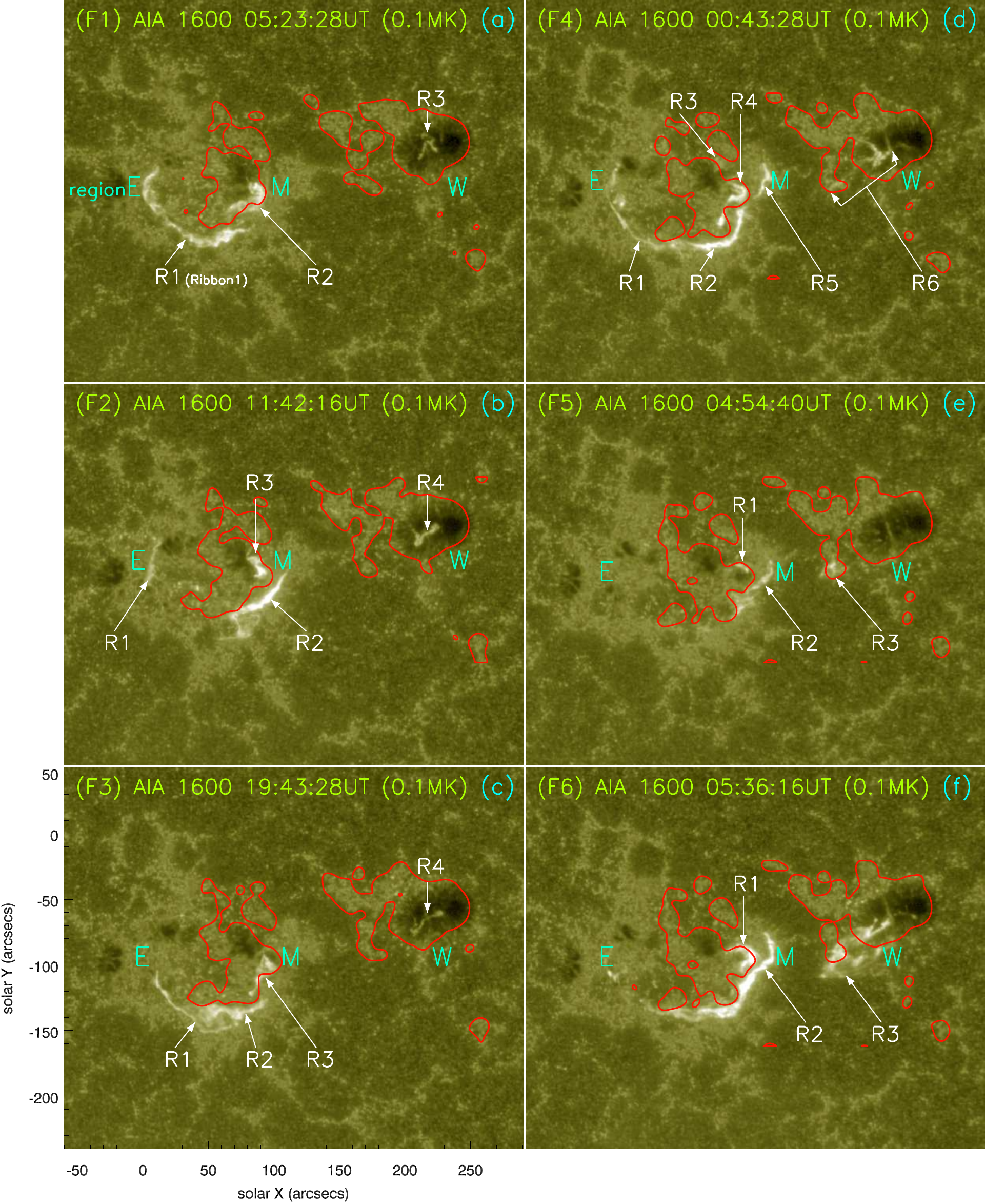}
      }
\caption{ \textbf{\textit{SDO}/AIA 1600 {\AA} images of the six flares (F1--F6).} The typical temperature for 1600 \AA\ emission is $\sim$ 0.1 MK. The red contour shows the positive line-of-sight magnetic field with a level of +60 G. The active region is divided into three parts: the east, the middle, and the west parts, which are denoted by regions E, M, and W, respectively. The arrows point to the flare ribbons, which are labeled as R1--R6.}
\label{fig:2}
\end{figure}
\begin{figure} 
      \vspace{-0.0\textwidth}    
      \centerline{\hspace*{0.00\textwidth}
      \includegraphics[width=1.0\textwidth,clip=]{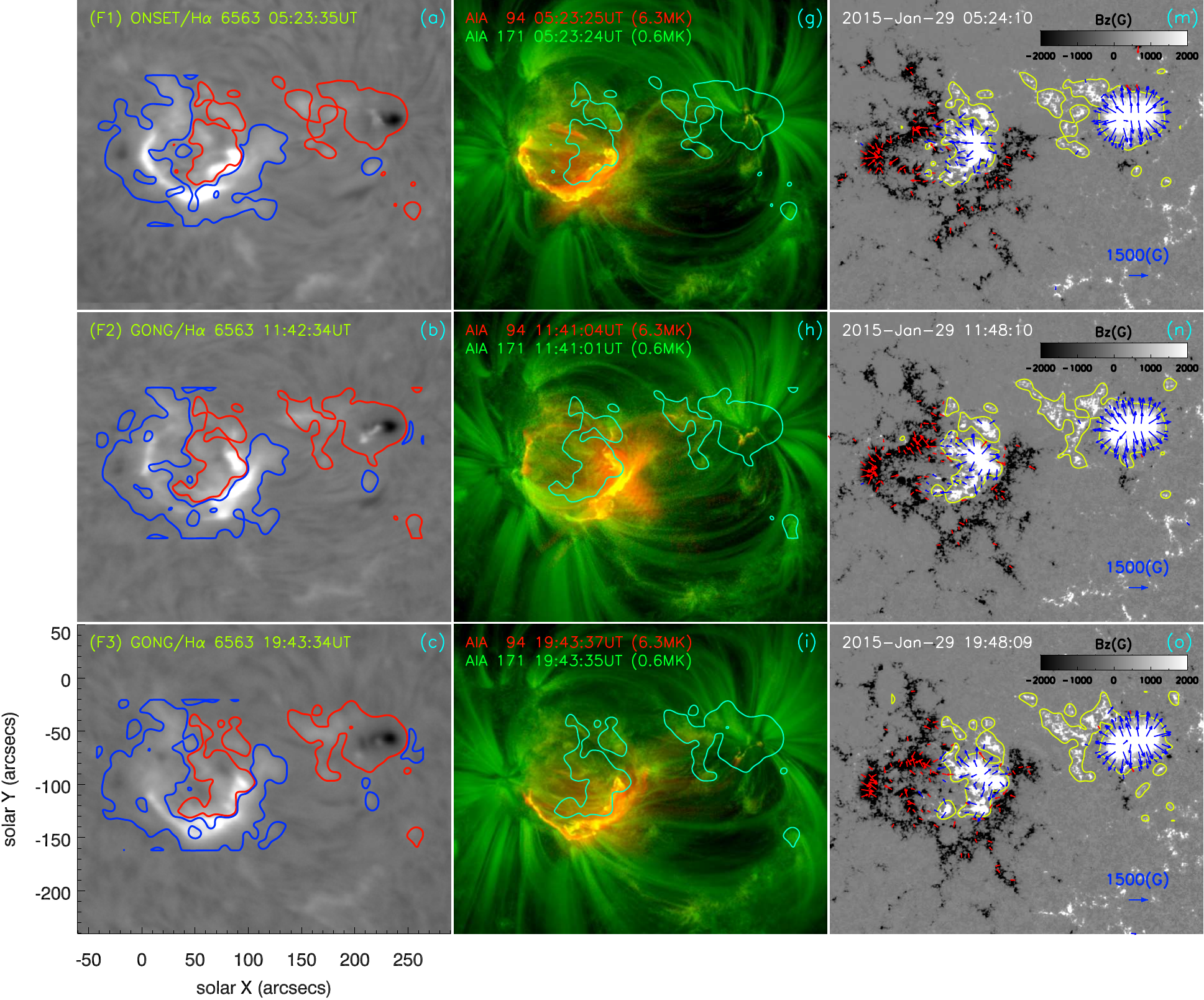}
      }
\caption{ \textbf{Overview of the evolution of the six flares (F1--F6).} Each row is for one flare. (a)--(f) H$\alpha$ 6563 {\AA} images observed by ONSET and GONG. The red and blue contours show the positive and negative line-of-sight magnetic field with levels of $\pm$ 60 G, respectively. (g)--(l) Composite images of \textit{SDO}/AIA 94 {\AA} (red) and 171 {\AA} (green), corresponding to emission temperatures of $\sim$ 6.3 MK and $\sim$ 0.6 MK, respectively. The cyan contour shows the positive line-of-sight magnetic field with a level of +60 G. (m)--(r) \textit{SDO}/HMI vector magnetic field after removing the 180$^\circ$ ambiguity of the transverse field and correcting the projection effect in the Cartesian coordinate system. The yellow contour shows the positive vertical component $B_z$ with a level of +60 G. The red and blue arrows represent the direction and magnitude of the horizontal component. The size of each panel is $\sim350\arcsec \times 290\arcsec$. The center of the field of view is located at [115\arcsec, -95\arcsec] in the helioprojective coordinate system, where the center of the solar disk is located at [0\arcsec, 0\arcsec].}
\label{fig:3}
\end{figure}
\begin{figure} 
      \vspace{-0.0\textwidth}    
      \centerline{\hspace*{0.00\textwidth}
      \includegraphics[width=1.0\textwidth,clip=]{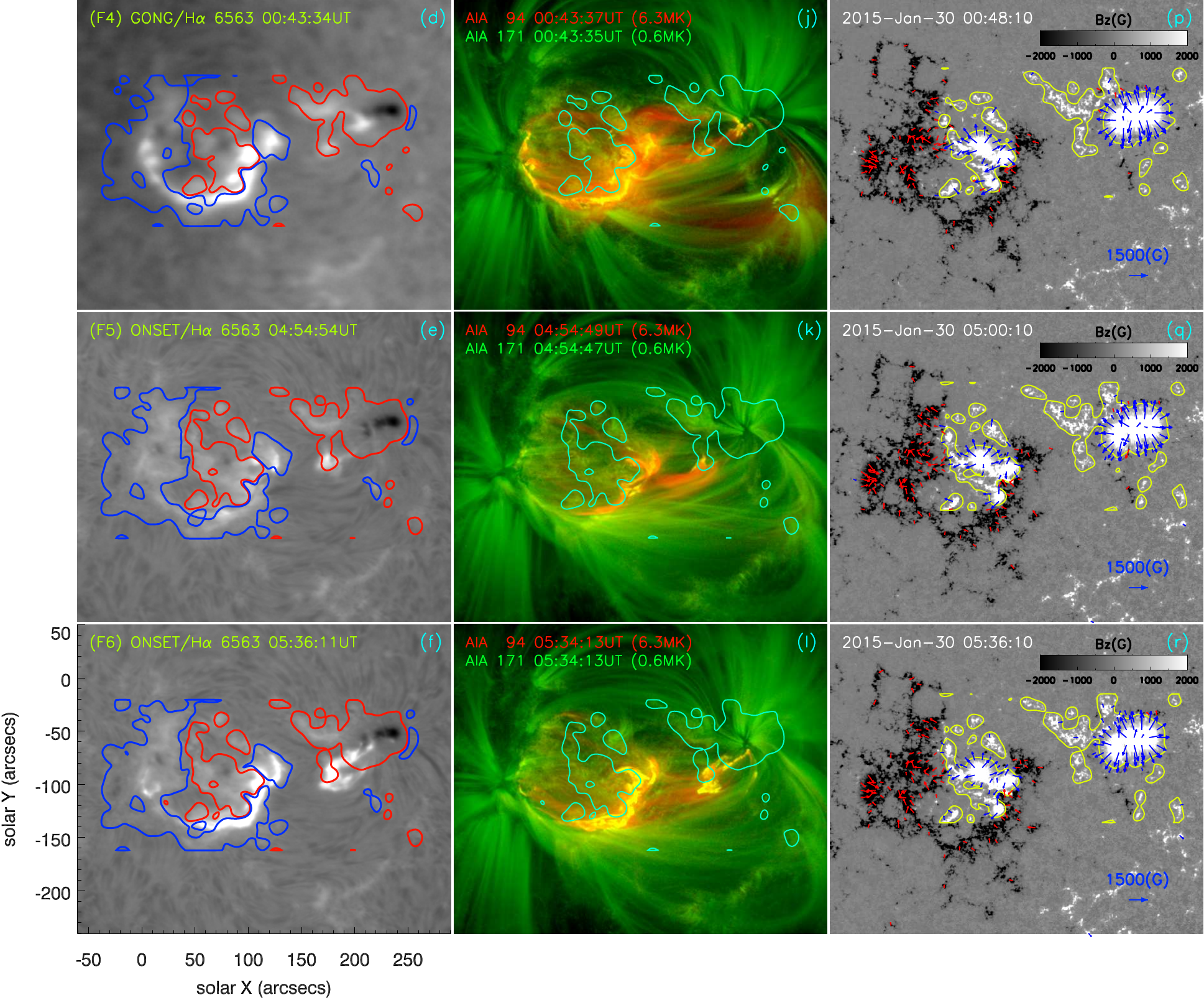}
      }
\figurenum{\ref{fig:3}}
\caption{Continued.}
\end{figure}
\begin{figure} 
      \vspace{-0.0\textwidth}    
      \centerline{\hspace*{0.00\textwidth}
      \includegraphics[width=1.0\textwidth,clip=]{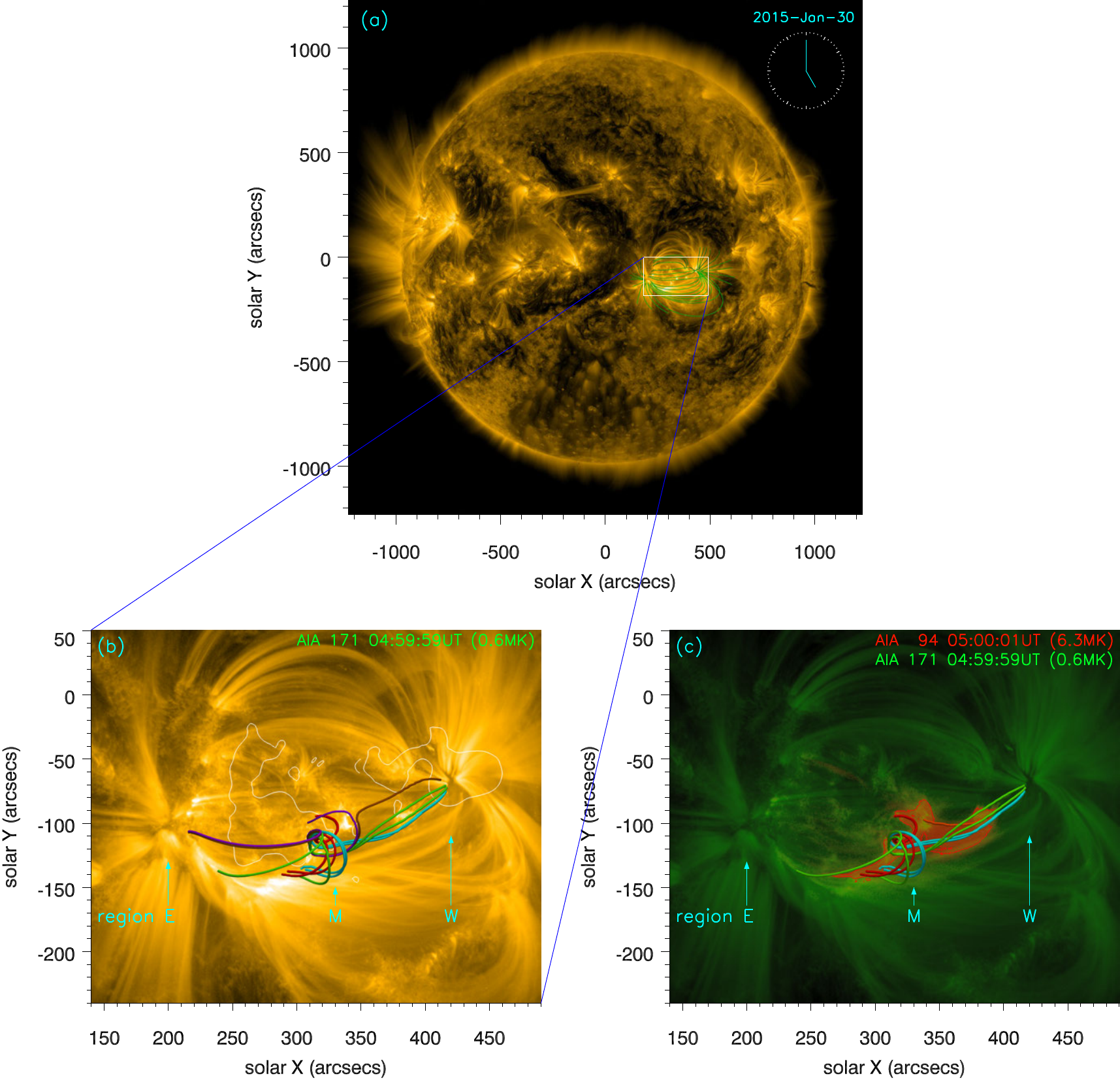}
      }
\caption{ \textbf{An example of the AIA extreme ultraviolet images and 3D coronal magnetic field lines at the moment of the flux rope bifurcation.} (a) Full disk image of \textit{SDO}/AIA 171 {\AA} at 04:59:59 UT on January 30. Overlaid are selected magnetic field lines at a relatively high altitude in the active region. (b) Zoomed-in view of the region corresponding to the white box in panel (a). Overlaid are selected magnetic field lines at a relatively low altitude. The cyan field lines connect regions M and W. The green and brown field lines link regions E and W. The purple field line links regions E and M. The red field lines show the twisted flux rope in region M. The white contour in the background shows the positive vertical component $B_z$ with a level of +40 G. (c) A composite image of \textit{SDO}/AIA 94 {\AA} (red) and 171 {\AA} (green) at 04:59:59 UT on January 29. The cyan, green, and red field lines are the same as that in panel (b). The red contour delineates the hot channel.}
\label{fig:4}
\end{figure}
\begin{figure} 
      \vspace{-0.0\textwidth}    
      \centerline{\hspace*{0.00\textwidth}
      \includegraphics[width=1.0\textwidth,clip=]{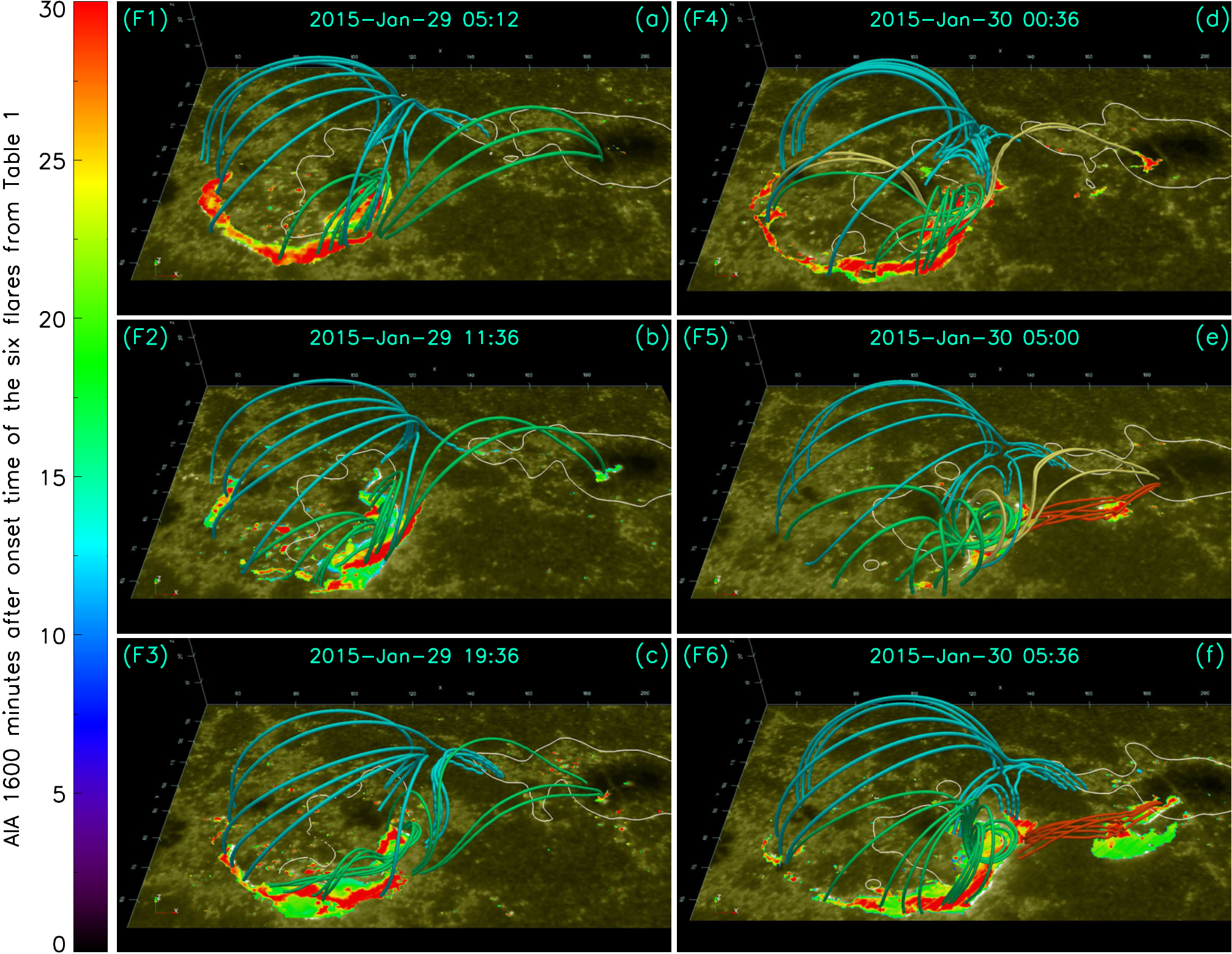}
      }
\caption{ \textbf{Selected 3D magnetic field lines in the corona and associated flare ribbons of the six flares (F1--F6).} The coronal magnetic field is reconstructed from photospheric vector magnetic field in the Cartesian coordinate at 05:24 UT, 11:36 UT, and 19:36 UT on January 29, and 00:36 UT, 05:00 UT, and 05:36 UT on January 30, respectively. The background shows the evolution of flare ribbons at \textit{SDO}/AIA 1600 {\AA} for 30 minutes from the onset time. The cyan lines show a large dome-like structure. The green lines include two parts, a part with low field lines connecting regions E and M, and the other part with high field lines linking regions M and W. The yellow lines extending from the western positive polarity become part of the magnetic flux rope at 05:00 UT on January 30. The red lines at a low altitude connect regions M and W. The white contour in the background shows the positive vertical component $B_z$ with a level of +40 G. The field of view (FOV) is $160 \times 85 \times 50$ grids with a uniform pixel size of 2\arcsec.}
\label{fig:5}
\end{figure}
\begin{deluxetable}{ccccccc}
\centering
\tablewidth{0pt}
\tablecolumns{7}
\tablecaption{List of null points of the six events in active region NOAA 12268 \label{t:1}}
\tablehead{
\colhead{Flare} & \colhead{Null point label} & \colhead{X position\tablenotemark{a} (Mm)} & \colhead{Y position\tablenotemark{a} (Mm)}   & \colhead{Height\tablenotemark{b} (Mm)}}
\startdata
F1 & Null point a           &      169.5 	& 161.4   &   8.0 \\
     & Null point b           &      170.5 	& 162.3   &   9.3 \\
F2 & Null point a           &      171.0 	& 146.9   &   3.3 \\
F3 & Null point a           &      175.6 	& 150.3   &   6.0 \\
     & Null point b           &      175.7 	& 153.7   &   8.7 \\ 
F4 & Null point a           &      169.2 	& 141.6   &   8.4 \\
     & Null point b           &      169.3 	& 141.5   &   8.8 \\     
F5 & Null point a           &      179.7 	& 143.4   &   8.4 \\
     & Null point b           &      181.4 	& 135.6   &   17.1 \\
     & Null point c           &      181.4 	& 135.6   &   17.2 \\
F6 & Null point a           &      179.0 	& 140.4   &   13.9 \\
     & Null point b           &      180.1 	& 136.7   &   17.9 \\
     & Null point c           &      181.7 	& 134.6   &   19.4 \\
     & Null point d           &      182.9 	& 131.9   &   20.1 \\
\enddata
\vspace{0.1in}
\tablenotemark{a}{The positions of X and Y are calculated from the east and south sides of the field of view, respectively.}\par
\vspace{0.05in}
\centerline{\tablenotemark{b}{The height of null points is measured from the Cartesian $z=0$ plane of our NLFFF computational box.}}
\end{deluxetable}
\begin{figure} 
      \vspace{-0.0\textwidth}    
      \centerline{\hspace*{0.00\textwidth}
      \includegraphics[width=1.0\textwidth,clip=]{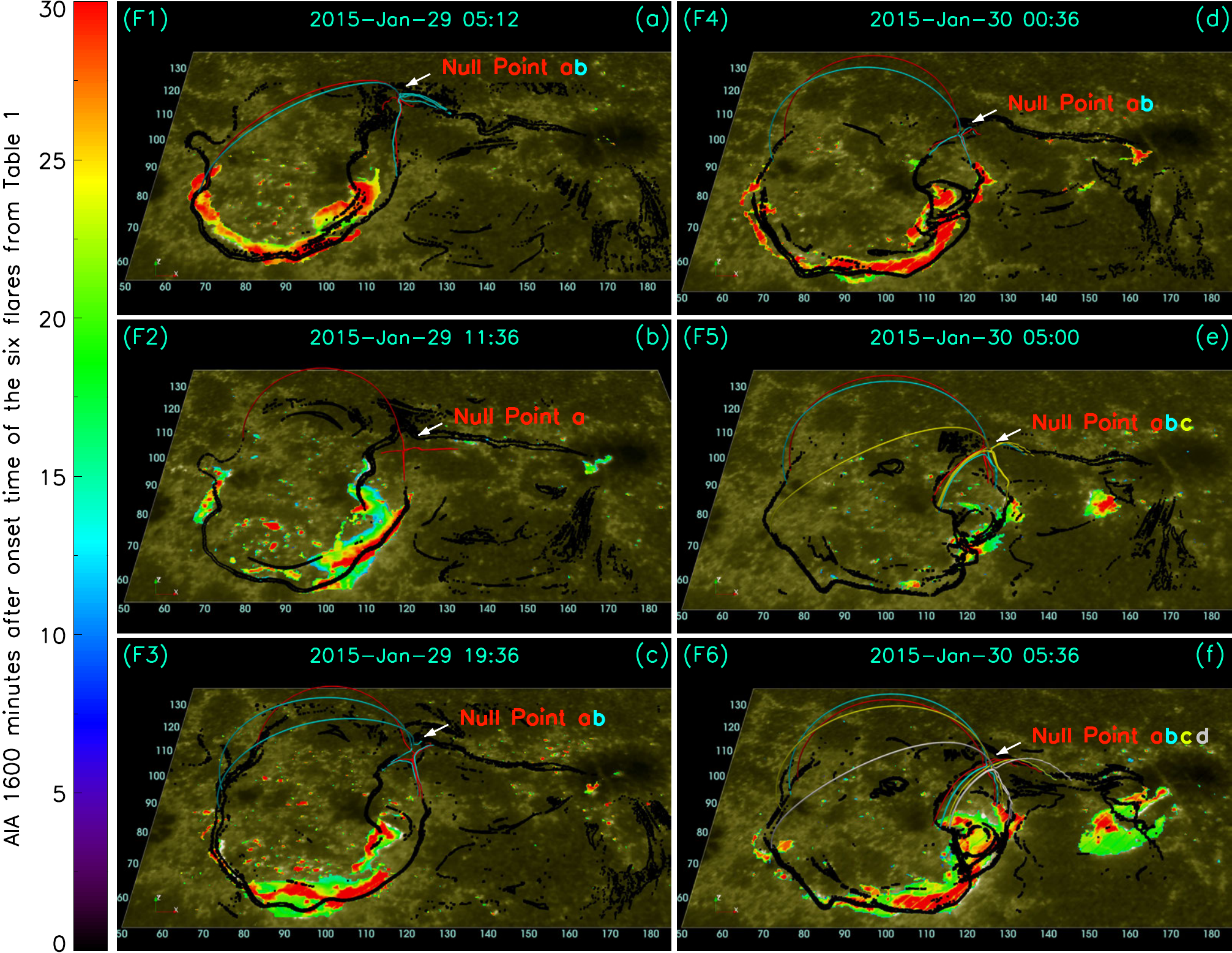}
      }
\caption{ \textbf{Magnetic field lines in the vicinity of magnetic null points for the six flares (F1--F6).} The white arrows point to the positions of null points. From low to high altitudes, the null points are labeled alphabetically and their associated field lines are represented by red, cyan, yellow, and gray lines, respectively. The background shows direct comparisons between QSLs at the $z=0$ layer with $Q$ values above $10^{3}$ (black solid lines) and the evolution of flare ribbons at \textit{SDO}/AIA 1600 \AA\ (same as the background of Figure 5).}
\label{fig:6}
\end{figure}
\begin{figure} 
      \vspace{-0.0\textwidth}    
      \centerline{\hspace*{0.00\textwidth}
      \includegraphics[width=1.0\textwidth,clip=]{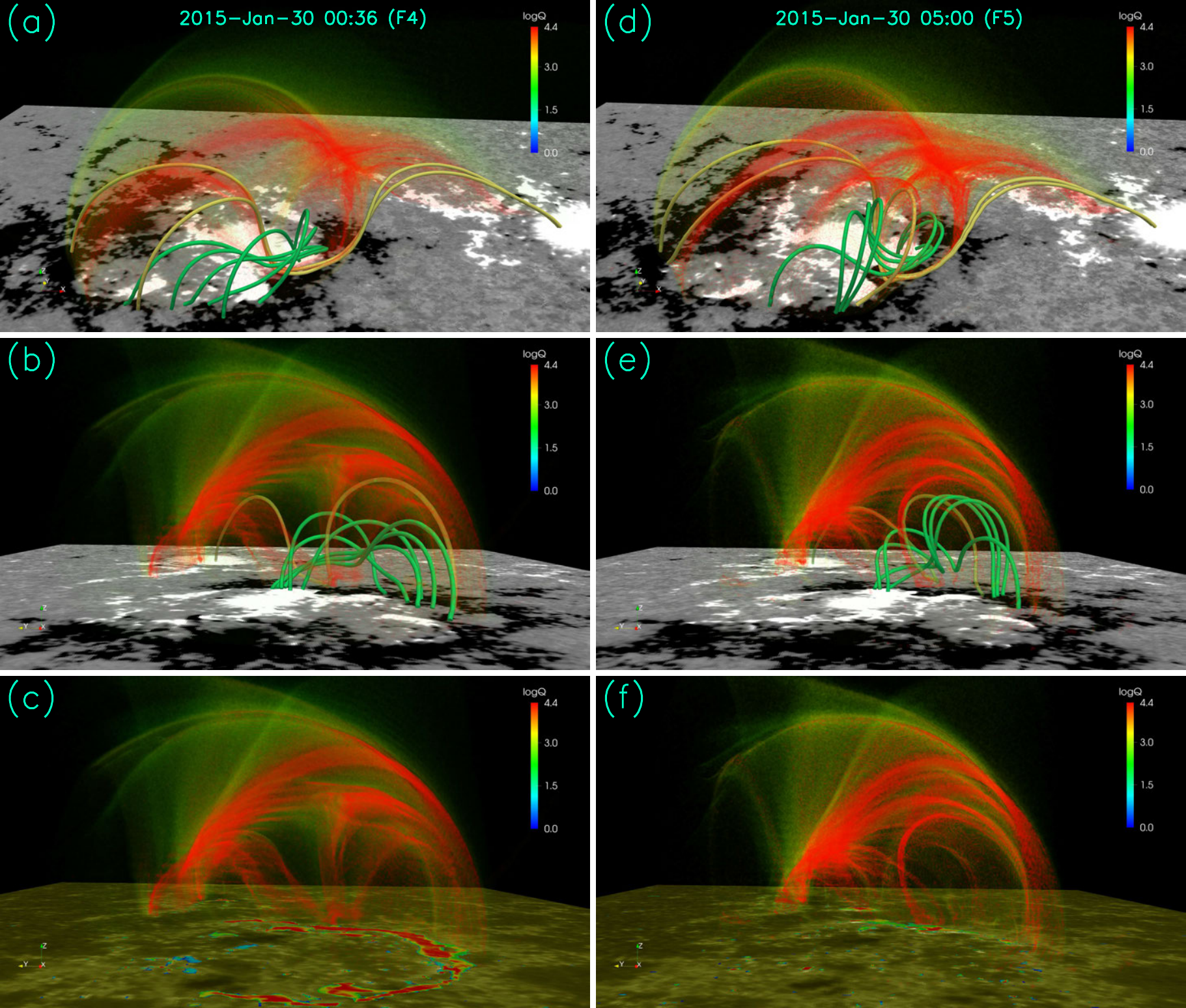}
      }
\caption{ \textbf{QSLs in flares F4 and F5 showing the topological structures at the flux rope bifurcation.} (a) A snapshot of 3D QSLs viewed along a selected direction for flare F4, which is calculated from the nonlinear force-free field. The yellow lines represent the bifurcated portion of the magnetic flux rope, while the green lines all lie within the large-scale dome-like QSL. The background image in black and white displays the \textit{SDO}/HMI $B_z$ component at 00:36 UT on January 30. (b)--(c) A snapshot of 3D QSLs viewed along the $X$-axis for flare F4. The field lines in panel (b) are the same as in panel (a). Panels (d)--(f) show the same quantities as panels (a)--(c), but for flare F5.}
\label{fig:6}
\end{figure}
\begin{figure} 
      \vspace{-0.0\textwidth}    
      \centerline{\hspace*{0.00\textwidth}
      \includegraphics[width=1.0\textwidth,clip=]{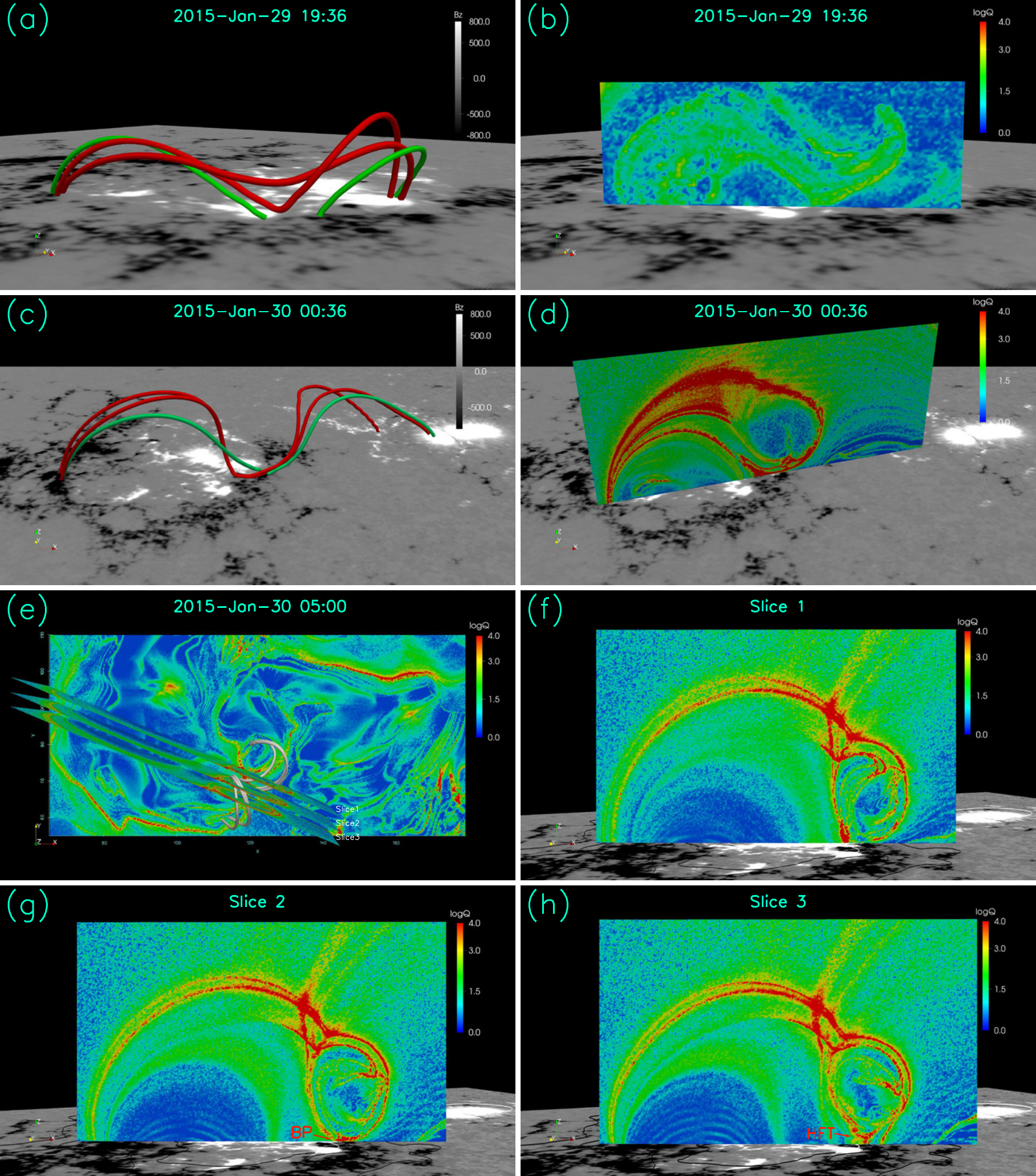}
      }
\caption{ \textbf{Magnetic topology at the moment of the magnetic flux rope formation and bifurcation.} (a) Bald-patch (BP) reconnection at the moment of magnetic flux rope formation. The green and red lines represent the magnetic field lines before and after magnetic reconnection, respectively. (b) $Q$-map slice roughly along the flux rope axis. The FOV is the same as panel (a). (c) BP reconnection at the moment of the magnetic flux rope bifurcation. The green and red lines have the same meaning as in panel (a). (d) $Q$-map slice roughly along the field lines. The FOV is the same as panel (c). (e) A $Q$-map on the bottom that is calculated from the nonlinear force-free field in the $xy$ plane at $z$ = 0. Overplotted are the three slices shown in panels (f)--(h). (f)--(h) $Q$-map slices perpendicular to the $xy$ plane at positions marked in panel (e). Panel (g) shows a BP structure and panel (h) displays an HFT structure at a low height. In all panels except panel (e), the background in black and white shows the \textit{SDO}/HMI $B_z$ component.}
\label{fig:7}
\end{figure}
\begin{figure} 
      \vspace{-0.0\textwidth}    
      \centerline{\hspace*{0.00\textwidth}
      \includegraphics[width=1.0\textwidth,clip=]{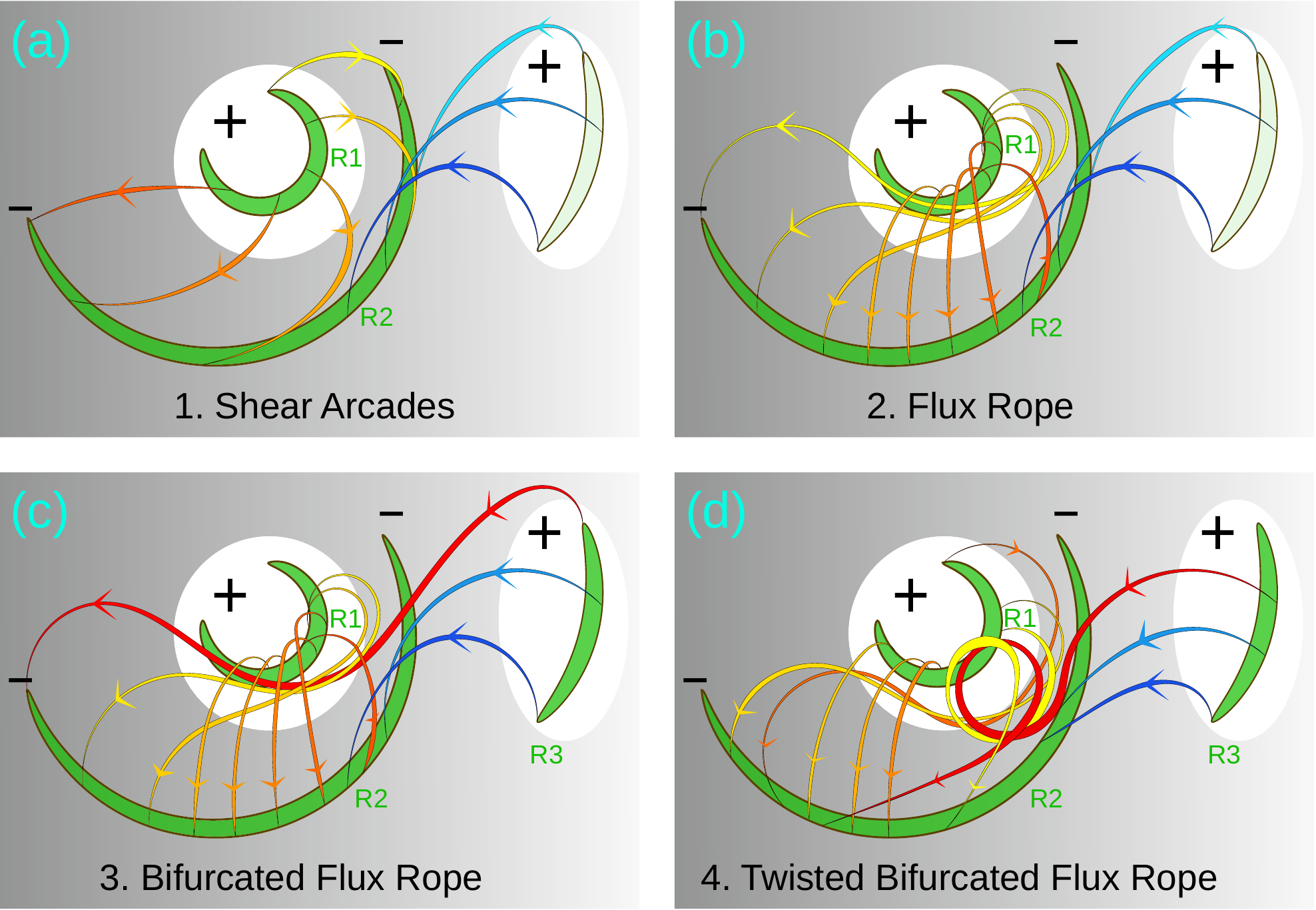}
      }
\caption{ \textbf{Schematic picture demonstrating the formation and bifurcation of the magnetic flux rope.} Green bands (R1, R2, and R3) represent flare ribbons. The warm color field lines show that the eastern positive polarity connects to the middle negative polarity. The cold color field lines display that the western positive polarity links to the middle negative polarity. The red field line is a part of the magnetic flux rope. The background includes two positive polarities and one negative polarity. (a) Shear arcades. (b) Flux rope. (c) Bifurcated flux rope. (d) Twisted bifurcated flux rope.}
\label{fig:8}
\end{figure}
\end{document}